\documentclass[twocolumn]{aastex631}

\usepackage{amsmath}
\usepackage{bm}

\newcommand{\rev}[1]{#1}

\accepted{March 3, 2025}

\shorttitle{Jet Archaeology and Forecasting}
\shortauthors{Tsunetoe et al.}

\begin{document}

\title{Jet Archaeology and Forecasting: Image Variability and Magnetic Field Configuration}

\correspondingauthor{Yuh Tsunetoe}
\email{ytsunetoe@fas.harvard.edu}

\author[0000-0003-0213-7628]{Yuh Tsunetoe}
\affiliation{Black Hole Initiative, Harvard University, \ 20 Garden street, Cambridge, \ MA 02138, USA}
\affiliation{Center for Computational Sciences, University of Tsukuba, \ 1-1-1 Tennodai, Tsukuba, \ Ibaraki 305-8577, Japan}

\author[0000-0002-1919-2730]{Ramesh Narayan}
\affiliation{Black Hole Initiative, Harvard University, \ 20 Garden street, Cambridge, \ MA 02138, USA}
\affiliation{Harvard-Smithsonian Center for Astrophysics, 60 Garden Street, Cambridge, MA 02138, USA}

\author[0000-0001-5287-0452]{Angelo Ricarte}
\affiliation{Black Hole Initiative, Harvard University, \ 20 Garden street, Cambridge, \ MA 02138, USA}
\affiliation{Harvard-Smithsonian Center for Astrophysics, 60 Garden Street, Cambridge, MA 02138, USA}



\begin{abstract}

We investigate how magnetic field variations around accreting black holes on event horizon scales affect the morphology of  magnetically-driven jet on larger scales. By performing radiative transfer calculations on general relativistic magnetohydrodynamics simulations, we find that temporal variation in the magnetic flux on the event horizon and the jet power are imprinted on the variability of jet width up to several hundred gravitational radii. When the magnetic flux around the black hole drops and then rises, the jet initially narrows or becomes truncated, then widens, creating a thin-thick pattern that propagates down the jet. This suggests that extended jet observations can provide a history record of horizon-scale magnetic field dynamics, and conversely,  upcoming changes in the jet image can be predicted from direct observation of the magnetized accreting plasma near the black hole. Furthermore, the pattern of jet width variations shows acceleration up to the relativistic regime as it moves away from the black hole, aligning with plasma bulk motion. We also find in time-averaged images that both the bulk plasma motion and magnetic field configuration in the jet-launching region, which are sensitive to black hole spin, shape diverse features through relativistic beaming and aberration. Higher black hole spins result in more poloidal bulk motion and toroidal magnetic fields, leading to more symmetric jet images and linear polarization patterns. These results suggest a new method for testing the magnetically arrested disk model and the Blandford-Znajek process, and for determining the black hole spin through observations bridging horizon and jet-launching scales. 

\end{abstract}



\section{Introduction}

In recent years, very long baseline interferometer (VLBI) observations by the Event Horizon Telescope (EHT) Collaboration have achieved direct imaging of supermassive black holes (BH) in the centers of nearby galaxies on event horizon scales (\citealp{2019ApJ...875L...1E,2022ApJ...930L..12E}; see also \citealp{2023Natur.616..686L}). 
M87, one of the primary targets, has a well-studied jet that has now been imaged from scales of sub-mpc to $\sim$80 kpc \citep[e.g.,][]{2012A&A...547A..56D,2016ApJ...817..131H,2018ApJ...855..128W,2023Natur.616..686L}. 
Upgrades and extensions to the EHT via the next-generation EHT (ngEHT) project and Black Hole Explorer (BHEX) will achieve at least a resolution of 5 $\mu$as (micro-arcseconds) at 230 GHz and a dynamic range close to 1000, from which movies connecting the inner accretion disk to the jet will be possible \citep{2023Galax..11..107D,2024arXiv240612917J}.  Thus, jet simulations providing detailed predictions at high-resolution and in the time domain are of timely importance.

So far, polarized EHT images favor strongly magnetized Magnetically Arrested Disk (MAD) models over their more weakly magnetized Standard And Normal Evolution (SANE) counterparts, the latter of which tend to be more weakly polarized and less consistent with multi-wavelength constraints \citep{2021ApJ...910L..13E,2019ApJ...875L...5E,2024ApJ...964L..26E}. 
In MAD models, the magnetic field advected along with the accreting plasma saturates and becomes dynamically important, leading to non-axisymmetric structures in the inflow and efficient launching of jets \citep{2003PASJ...55L..69N,2003ApJ...592.1042I}.  MADs also exhibit disruptive magnetic reconnection or ``flux eruption events,'' which offer an electron acceleration mechanism hypothesized to explain multi-wavelength polarized flares \citep[e.g.,][]{2020MNRAS.497.4999D,2022ApJ...924L..32R,2022ApJ...941...30C,2024MNRAS.531.3961N}.
MAD models accelerate efficient relativistic jets powered by the spin-energy of the BH, tapped by magnetic fields threading the horizon (\citealp{1977MNRAS.179..433B}; BZ process; see also \citealp{2011MNRAS.418L..79T}). It will be important to verify predictions of MAD models with observations in the jet launching region as well.

Previous observational studies have explored variable features of parsec-scale jets after X- and gamma-ray flares (for example, light curve; \citealp{2010ApJ...722L...7P}; relativistic blobs; \citealp{2001ApJ...556..738J,2013ApJ...773..147J}; jet orientation; \citealp{2014A&A...571L...2R}; and jet base; \citealp{2015ApJ...807L..14N,2017MNRAS.468.4478L}; \rev{see also \citealp{2021MNRAS.502.2023P} for infrared flares in the Galactic Center}). These high-energy flares can be associated with eruptive events in the vicinity of the BH (\citealp{2010MNRAS.405L..94T,2014A&A...571L...2R,2015ApJ...809...97B}; see also \citealp{2012A&A...537A..70S,2014ApJ...780...87M} for another possibility of the flares taking place far from the BH). 
Most recently, \cite{2024arXiv240417623T} reported changes in the position angle of the M87 jet orientation andthe bright spot on the M87* ring at the same time with a gamma-ray flare in 2018, though the emission origin of the flare and its relation to the jet orientation are uncertain. 

\bigskip

Polarization images can also be a strong tool to address the magnetically-driven jet mechanism and have attracted a large interest. An ordered linear polarization (LP) vector pattern has been detected around the photon ring (\citealp{2021ApJ...910L..12E}) and the extended jet (e.g., \citealp{2018ApJ...855..128W}), which implies the existence of a persistent magnetic field structure bridging these scales. 

In theoretical approaches, polarization images have been suggested as a tool for inferring black hole spin \citep{2020ApJ...894..156P,2023ApJ...958...65C,2023ApJ...950...38E,2023MNRAS.520.4867Q,2023Galax..11....6R,2024ApJ...964L..26E}, although interpreting these images can be challenging due to Faraday effects, particularly Faraday rotation. 
Studies have also investigated the Faraday effects on the polarization images around the BH (\citealp{2017MNRAS.468.2214M,2020MNRAS.498.5468R,2022ApJ...931...25T}), as well as relativistic effects from
accelerated plasma bulk motion (\citealp{1981ApJ...248...87L,2005MNRAS.360..869L,2010ApJ...725..750B,2013MNRAS.430.1504M}). 
It has also been suggested that circular polarization (CP) images can shed new light on the jet mechanism  (\citealp{2020PASJ...72...32T,2021PASJ...73..912T,2021MNRAS.505..523R,2021MNRAS.508.4282M,2023ApJ...957L..20E,2024ApJ...972..135J}). 

On the scale of more extended jets, the effect of magnetic field helicity combined with viewing angle was examined in \cite{1981ApJ...248...87L} and \cite{2013MNRAS.430.1504M} using one-zone cylindrical jet models. 
\cite{2021A&A...656A.143K} calculated polarization images based on relativistic MHD jet models with different magnetic field configurations, obtaining a bimodal pattern of horizontal and vertical LP vectors to the jet (see also \citealp{2003A&A...403..805P,2011ApJ...737...42P}). 
\cite{2023ApJ...959L...3D} also investigated the effect of waves on shearing flow surfaces on the LP features. 

\bigskip

Motivated by this previous activity and by upcoming observations, we examine the relationship between the horizon-scale dynamical variability in MAD simulations and the polarized morphology of the jet launching region in synthetic images. 
\cite{2022MNRAS.511.3795N} surveyed the jet and disk morphology in MAD GRMHD simulations for a variety of BH spins and found that the prograde spin cases give wider jets up to $100~r_{\rm g}$ scale than the retrograde cases (see subsection \ref{subsec:GRMHD} for detail). 
Here $r_{\rm g} = GM_\bullet/c^2$, $G$ is the the gravitational constant, $M_\bullet$ is the BH mass, and $c$ is the speed of light. 
\cite{2023arXiv231100432C} also pointed out that the jet opening and tilt angles at $10~r_{\rm g}$ are correlated and anti-correlated with the normalized magnetic flux, respectively. 
\rev{
Recently, \cite{2024ApJ...960..106M} has also started surveying the jet image variability and the spin-dependence. 
}

Here, we focus on the transverse structure of the inner jet image for an M87-like system on a scale $\lesssim 100~r_{\rm g}$ ($\sim 0.003~{\rm pc} \sim 0.37~{\rm mas}$; milli-arcseconds), which corresponds to $\lesssim 300~r_{\rm g}$ if de-projected with the observer's inclination angle of $i = 163\degr$ (or $17\degr$). 
Since the plasma on these scales is being accelerated up to the relativistic regime, the jet morphology in the images is affected by relativistic beaming and aberration effects on the synchrotron radiation. 
With this in mind, we perform polarimetric general relativistic radiative transfer (GRRT) calculations based on MAD GRMHD models, to obtain theoretical prediction of images and investigate the relationship between the plasma dynamics and  observable image features. 

\bigskip

This paper is organized as follows. Our methodology is introduced in section \ref{sec:model}, which consists of GRMHD simulations (subsection \ref{subsec:GRMHD}) and GRRT calculations (subsection \ref{subsec:GRRT}). 
We describe our main results in section \ref{sec:result}, regarding the variability of jet shape in subsections \ref{subsec:snapshots} and \ref{subsec:correlation} and the averaged image features in subsections \ref{subsec:average}.
Applications of our results are addressed in section \ref{sec:discussion}, including an estimation of the jet acceleration profile in subsection \ref{subsec:acceleration}, the variability of the jet width in \ref{subsec:deviation}, correlation between the jet power and width in \ref{subsec:power-width}, the spin-dependence of polarization images in \ref{subsec:LPCPspindep}, and comparison with observations and contribution of nonthermal electrons in \ref{subsec:nonthermal}. 
Finally, we present a summary and conclusions in section \ref{sec:conclusions}.

\section{Method}\label{sec:model}

\subsection{MAD GRMHD Models}\label{subsec:GRMHD}

We use as our starting point the GRMHD simulations in \cite{2022MNRAS.511.3795N}, in which the authors surveyed MAD models for 9 BH spin values: $a_* \equiv a/M_\bullet \in \{0, \pm 0.3, \pm 0.5, \pm 0.7, \pm 0.9\}$. 
They found that the prograde spin models ($a_* > 0$) had wider jets and thinner disks than the retrograde models ($a_* < 0$). In addition, they also pointed out that models with larger absolute spin values $|a_*|$ exhibited more variability in $\phi_{\rm BH}$, the normalized magnetic flux threaded on the event horizon $\phi_{\rm BH}$. 
Here, 
\begin{equation}\label{eq:phi_BH}
    \phi_{\rm BH} \equiv \frac{\Phi_{\rm BH}}{\sqrt{\dot{M}r_{\rm g}^2c}},
\end{equation}
\begin{equation}
    \Phi_{\rm BH} = \frac{1}{2} \iint \sqrt{-g} |B^r|_{r = r_{\rm H}} {\rm d}\theta {\rm d}\phi,
\end{equation}
\noindent where $g$ is the determinant of the metric, $B^r$ is the radial component of the three-vector magnetic field, and $r_{\rm H}$ is the radius of the event horizon. 

Here, we select a subset of 5 BH spins, $a_* \in \{ \pm0.9, \pm0.5, 0\}$. For each model, we ray-trace snapshots separated by $50~t_{\rm g}$ over a total duration of $5000~t_{\rm g}$ in quasi-steady state, where $t_{\rm g} = r_{\rm g}/c$. 
The duration of $5000~t_{\rm g}$ corresponds to about five years in M87*, assuming $M_\bullet = 6.5\times10^9~M_\odot$ (e.g., \citealp{2019ApJ...875L...5E}).

To determine the electron temperature, we implement the R-$\beta$ prescription, 
\begin{equation}\label{eq:R-beta}
	\frac{T_{\rm i}}{T_{\rm e}} = R_{\rm low}\frac{1}{1+\beta^2} + R_{\rm high}\frac{\beta^2}{1+\beta^2},
\end{equation}
where $T_{\rm i}$ and $T_{\rm e}$ are proton and electron temperatures, respectively, and $\beta$ is the thermal-to-magnetic pressure ratio \citep{2016A&A...586A..38M}. 
Here, the two parameters are set to $R_{\rm low} = 1$ and $R_{\rm high} = 160$, where we choose the latter because polarized studies of M87* prefer large values \citep{2023ApJ...957L..20E}. 

\subsection{General Relativistic Radiative Transfer}\label{subsec:GRRT}

We ray-traced each GRMHD model snapshot using our polarimetric GRRT code \texttt{SHAKO} to calculate a series of images, or a movie (see \citealp{TsuY:2023} for the details of code composition. See also Appendix of \citealp{2024PASJ..tmp...84T} for comparison with another GRRT code \texttt{ipole} \citep{2018MNRAS.475...43M,2022ApJS..259...64W}). It is well-known that assuming thermal electron distribution functions is sufficient to explain the emission from the inner-most accretion flow, but non-thermal electrons are required to reproduce extended jets \citep[e.g.,][]{2019A&A...632A...2D,2022NatAs...6..103C,2022A&A...660A.107F}.  Therefore, we assume a combination of thermal and non-thermal synchrotron-emitting electrons, the latter of which are distributed in a power-law (see \citealp{TsuY:2023} and \citealp{2024PASJ..tmp...84T} for the radiative coefficients for thermal and power-law synchrotron emission, respectively).

The electron temperature of the thermal component is given by equation \ref{eq:R-beta}. The power-law index and minimum and maximum Lorentz factors of the nonthermal electrons are fixed to $p=2.5$, $\gamma_{\rm min} = 30$ and $\gamma_{\rm max} = 10^6$ in the whole region. The energy partition between thermal and power-law electrons, $u_{\rm e,pl} = 0.03~u_{\rm e,th}$, is assumed to determine the number density of power-law electrons (\citealp{2003ApJ...598..301Y,2012MNRAS.421.1517D}; see also subsection \ref{subsec:nonthermal}). 

We set the observer's screen at a distance of $r = 10^{4}~r_{\rm g}$ with an inclination angle of $i = 163\degr$ ($17\degr$) to the $z$-axis for the prograde (retrograde) spin cases. 
The radiative transfer is performed under the assumption of fast-light, in which the plasma is stationary during the propagation of light rays. 
We cut out the region with the magnetization $\sigma_{\rm m} \equiv b^2/4\pi\rho c^2 >10$ to avoid regions where the density floor, used in the GRMHD code to maintain numerical stability, artificially injects material. 
Regions beyond $R = r~{\rm sin}\theta > 200~r_{\rm g}$ are also cut out because the accretion flow there may not be in equilibrium. 
We scale the particle density and magnetic fields from the scale-invariant ideal GRMHD simulation, to reproduce an average flux of $\sim 1~{\rm Jy}$ at 230~GHz for each model.  This is larger by a factor of 2 than the value of 0.5 Jy adopted for most EHT studies \citep[e.g.,][]{2019ApJ...875L...5E}, since we now include emission from the jet region, not just the ``compact'' shadow region around the black hole.

In the following, we show the 86 GHz images with $260 \times 520$ pixels, convolved with a circular Gaussian beam with a FWHM (full width of half maximum) of  $40~\mu{\rm as}$ \citep[similar resolution to][]{2023Natur.616..686L},\footnote{We do not see a ring feature in the innermost region of the convolved images, while the observational images in \cite{2023Natur.616..686L} exhibits a ring with a diameter of $\approx 64~{\rm \mu as}$. This is because the size of the photon ring, $\approx 10~r_{\rm g} \approx 37~{\rm \mu as}$, is smaller than the beam size $40~\mu{\rm as}$ and remains unresolved. \cite{2024ApJ...969...22O} obtained a $\sim 60~{\rm \mu as}$ ring using a semianalytic funnel jet model, which consists of emission from the stagnation surface between the inflow and outflow plasmas in the funnel region.} unless otherwise noted.

\section{Results}\label{sec:result}

\subsection{Truncation of Jet Associated with a Deficit of Magnetic Flux}\label{subsec:snapshots}

\begin{figure*}
\begin{center}
	\includegraphics[width=19cm]{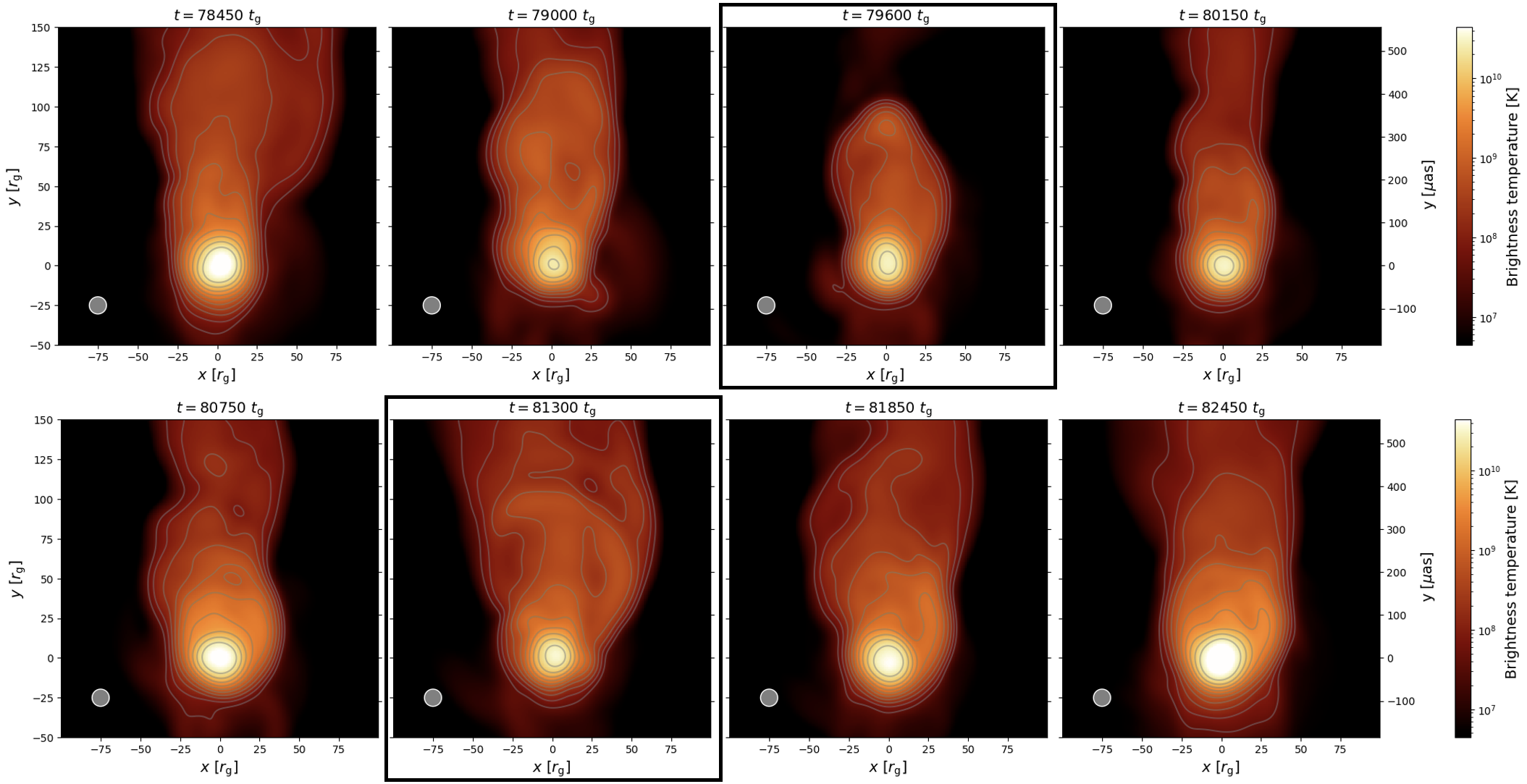}
    \includegraphics[width=12cm]{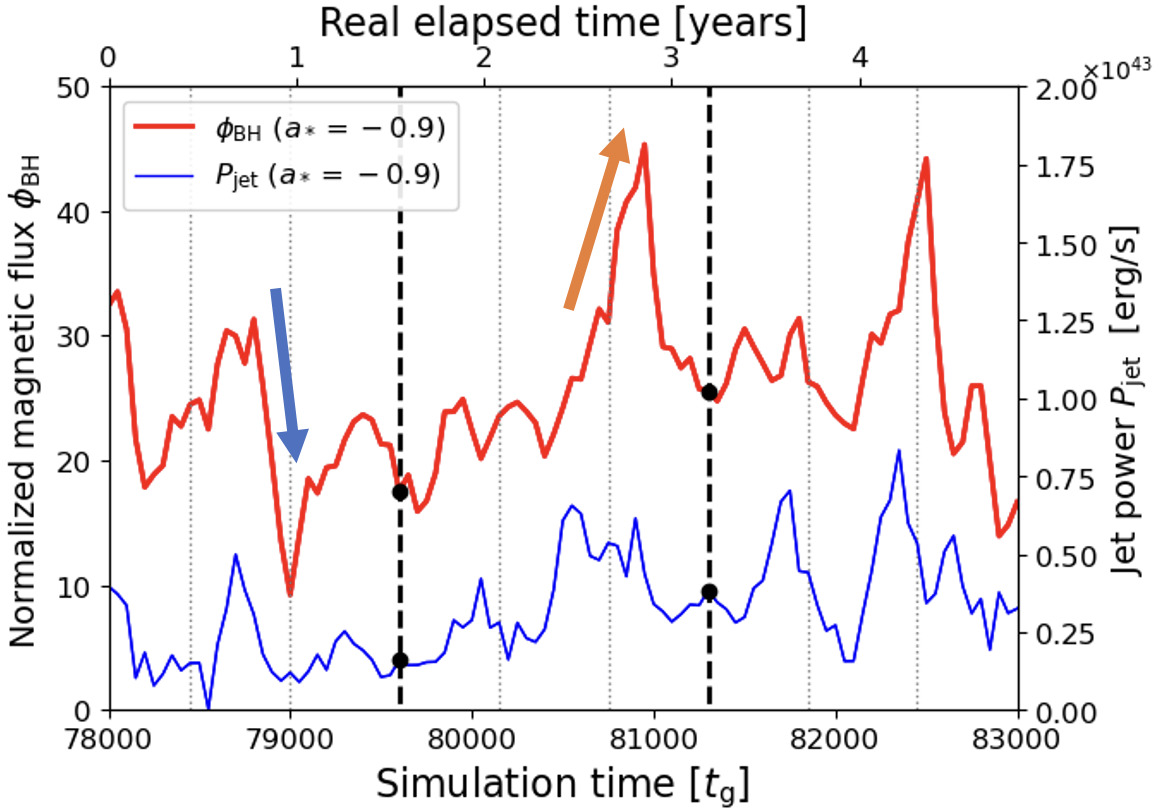}
\end{center}
    \caption{
    Top: eight snapshot images of the total intensity at 86~GHz for a GRMHD model with $a_* = -0.9$. 
    The projection of the BH spin axis and the orientation of the approaching jet on the sky both point upward on the images. 
    The contour levels increase in steps of 2. 
    The gray circle in the bottom left of the images shows the size of the FWHM of the circular Gaussian convolution beam.
    Bottom: the profiles of the normalized magnetic flux on the event horizon $\phi_{\rm BH}$ and the jet power $P_{\rm jet}$ for the $a_* = -0.9$ model as a function of time. The two bold dashed lines and six dotted lines correspond to the times of the 8 snapshots at the top. The blue and orange arrows indicate a drop and rise in $\phi_{\rm BH}$, respectively, prior to the two snapshots at $t = 79600~t_{\rm g}$ and $81300~t_{\rm g}$. 
    See \url{https://youtu.be/FwiKspIXjZk} for a movie.
    }
    \label{fig:snapshots}
\end{figure*}


We show eight total intensity snapshot images for the $a_* = -0.9$ model in Fig.~\ref{fig:snapshots}. Although $a_* = -0.9$ may be intrinsically unlikely \citep[e.g.,][]{2024arXiv241007477R}, we showcase this model due to an illustrative flux eruption event we identified. 
They exhibit a large variability in jet morphology, reflecting large fluctuations in the magnetic field in the underlying in MAD model.
In particular, we focus on the two squared images at $t = 79600~t_{\rm g}$ and $81300~t_{\rm g}$. 
These two images exhibit a truncation and broadening of the approaching jet, exemplifying the dramatic temporal variability possible in MAD models. 

Here, we focus on the variability of the normalized magnetic flux on the event horizon, $\phi_{\rm BH}$ defined in Eq.~\ref{eq:phi_BH}, as shown in the profile in Fig.~\ref{fig:snapshots}. 
In addition, we plot \rev{the jet power $P_{\rm jet}$, the outflowing energy flux estimated from GRMHD data}. 
The two bold dashed lines in the bottom panel of Fig.~\ref{fig:snapshots} denote the times of the squared snapshot images at the top. 
During the time period marked with a blue arrow, $\phi_\mathrm{BH}$ sharply drops, corresponding to a magnetic reconnection or ``flux eruption'' event.  Since the jet power follows $P_\mathrm{jet} \propto \phi_{\rm BH}^2$ (\citealp{1977MNRAS.179..433B,2010ApJ...711...50T}), the jet temporarily loses power.  The ``truncated'' jet in the top left panel thus represents a shorter newly recovered jet.  In the time period marked by the orange arrow, the magnetic flux rapidly increases, leading to an increase in jet power.  As a result, the jet profile becomes wider.  At the slightly later time in the second dashed line, the jet has returned to a more moderate width as in the seventh panel at $t = 81850~t_{\rm g}$.  

The above comparison shows that variations in $\phi_\mathrm{BH}$ lead to variations in the jet width as a function of distance, due to causal variations in the jet power.  In this way, the temporal history of $\phi_\mathrm{BH}$ is encoded in the {\it spatial} variation of the jet width.\footnote{In observations, \cite{2016ApJ...817..131H} reported a ``constricted'' structure at $\sim 0.2-0.3~{\rm mas}$ in M87 jet. }

\subsection{Correlation of Jet Widths and Magnetic Flux}\label{subsec:correlation}

\begin{figure*}
\begin{center}
	\includegraphics[width=15cm]{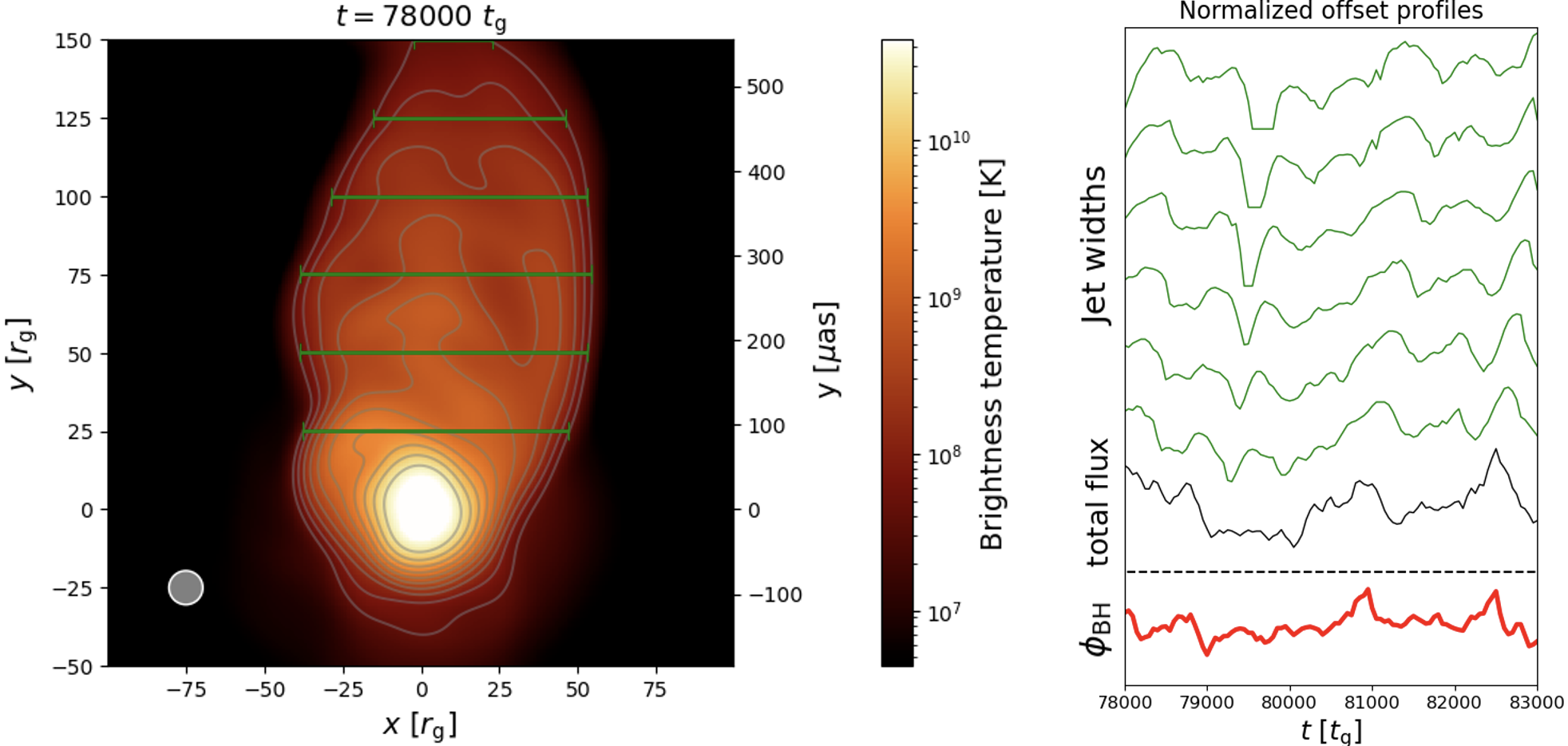}
\end{center}
    \caption{
    Left: Total intensity snapshot at time $t = 78000~t_{\rm g}$ for the $a_* = -0.9$ model where our calculated jet widths are visualized using green lines, at six projected altitudes $y = 25r_{\rm g}, \ 50r_{\rm g}, \ 75r_{\rm g}, \ 100r_{\rm g}, \ 125r_{\rm g}$, and $150r_{\rm g}$. Here we define the two edges of the jet as the outermost points in the transverse profile at which the intensity is greater than $10^{-3}$ times the peak color contour.
    Right: Variability profiles of $\phi_{\rm BH}$ (red), the total flux (black), and the six jet widths (green) defined as in the left for a duration of $5000~t_{\rm g}$. The absolute scales of the profiles are ignored here. See \url{https://youtu.be/FwiKspIXjZk} for a movie.
    }
    \label{fig:profiles}
\end{figure*}


To verify the hypothesis of $\phi_{\rm BH}$-jet width correlationship, we introduce a measure of jet width at altitudes of $y = 25, 50, 75, 100, 125,$ and $150~r_{\rm g}$ on the image.\footnote{Note that because of the inclination, these $y$ distances correspond to real physical distances of
$85, 170, 255, 340, 425,$ and $ 510 r_{\rm g}$ from the BH, respectively.}
Here, we measure the width as the offset between the two edges, defined as the leftmost and rightmost pixels on the transverse profile where the total intensity exceeds $10^{-3}$ times the peak intensity. 
We set the peak intensity to $1 \times 10^{-4}~{\rm [erg~s^{-1}~cm^{-2}~Hz^{-1}]} \approx 4.4\times10^{10}~[{\rm K}]$ in the brightness temperature, based on an average intensity over $5000~t_{\rm g}$. 
The jet widths for a snapshot are denoted by the green horizontal lines in the left panel of Fig.~\ref{fig:profiles}. 
In the right panel, we plot jet width as a function of time, along with time variation of $\phi_{\rm BH}$ and the total flux density at 86~GHz, which is obtained by integrating the total intensity over the image. Here we disregard the absolute scales of the profiles and offset them for illustrative purposes. 

We can see at a glance that the jet widths exhibit similar temporal variability to both $\phi_{\rm BH}$ and the total flux, but with a time lag. As mentioned in the previous subsection, it is confirmed that the large drop in $\phi_{\rm BH}$ around $t = 79000t_{\rm g}$ is followed by a series of jet narrowing or truncation from downstream (small $y$) to upstream (large $y$), 
while the rise around $t = 81000t_{\rm g}$ is followed by a horizontal expansion of the jet. 
The total flux also tends to show a similar variability to $\phi_{\rm BH}$, with a shorter time lag.
This result implies that there is a strong connection between the horizon-scale dynamics and the extended jet image at scales hundreds of times larger.
Thus, high resolution imaging would enable ``jet archaeology'' by allowing us to infer the temporal variation in $\phi_\mathrm{BH}$ from the spatial variation in width.
At the same time, light curves would enable ``jet forecasting'' by allowing us to predict future jet width profiles from variations in total flux.

\bigskip

\begin{figure*}
\begin{center}
	\includegraphics[width=16cm]{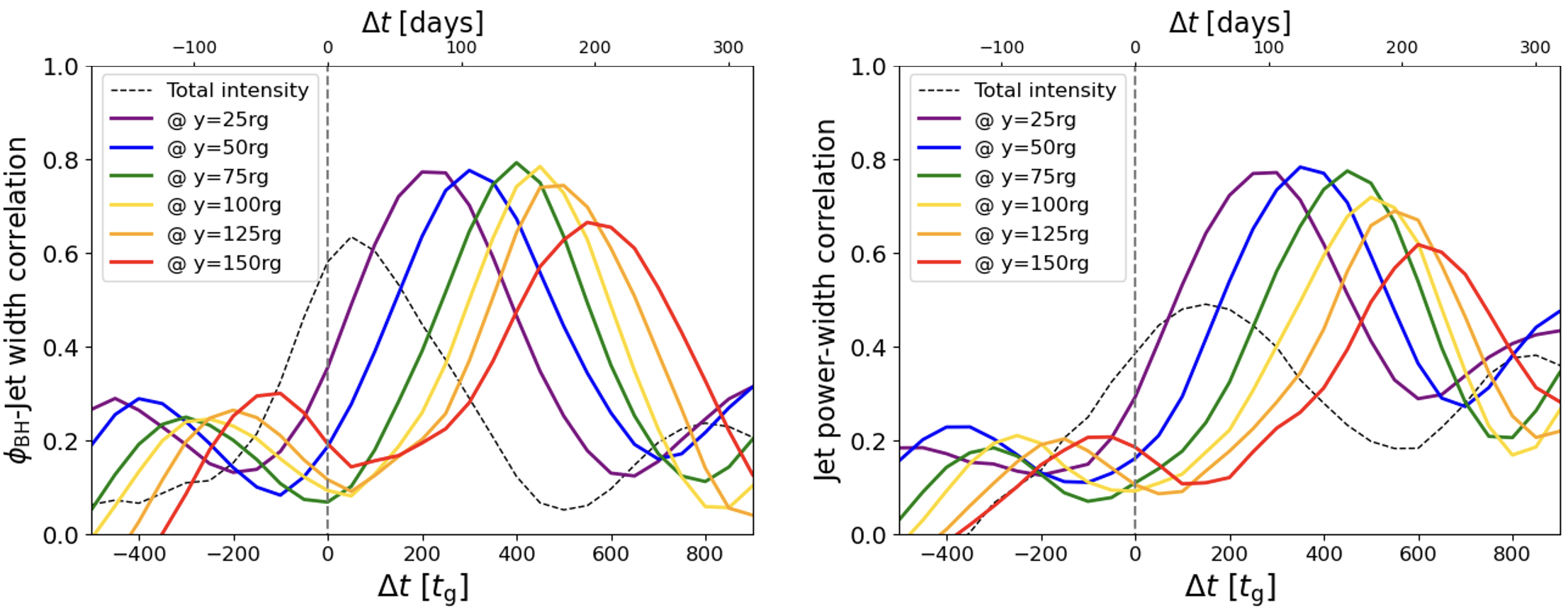}
\end{center}
    \caption{
    Left: correlation functions between $\phi_{\rm BH}$ and the jet widths (solid lines), along with that between $\phi_{\rm BH}$ and total flux (dashed line). 
    Right: correlation functions between the jet power $P_{\rm jet}$ and jet widths.
    }
    \label{fig:correlation}
\end{figure*}

To evaluate the strength of these correlated variations, we introduce the correlation function
\begin{equation}
    {\rm corr}(\Delta t) \equiv \int_{t_{\rm ini}}^{t_{\rm fin}} {\rm d}t 
        ~\overline{w_{\rm jet}}(t+\Delta t) ~\overline{\phi_{\rm BH}}(t),
\end{equation}
where ${w_{\rm jet}}(t)$ is the jet width at time $t$ and the bar over the variables denotes normalization over a duration of $5000~t_{\rm g}$. The obtained correlations at each altitude on the image are plotted in the left panel of Fig.~\ref{fig:correlation}, along with the correlation between $\phi_{\rm BH}$ and the total flux. 

We find that the jet widths show a stronger correlation with $\phi_{\rm BH}$, peaking at around 0.8, than the total flux.
This means that while the total flux is a good indicator, dominated by radiation from the photon ring and strongly reflecting the magnetic flux, the jet widths serve as an even better indicator of the variability in $\phi_{\rm BH}$. 
The $\Delta t_{\rm peak}$, time lag, is larger for higher $y$, as evident already in Fig.~\ref{fig:profiles}.

In the above, we saw the strong connection between the horizon-scale dynamics and the extended jet shape for the $a_*=-0.9$ model.\footnote{\rev{We also confirm in Appendix \ref{apdx:survey} that the correlation holds for various model parameters.}} 
In Appendix \ref{apdx:a09}, we show qualitatively similar results for the other spinning BH models $a_* = +0.9, \pm0.5$ (see the next subsection for the absence of jet in the non-spinning case).
It is confirmed for the four spinning models that the jet widths strongly reflect the $\phi$-variability as long as both edges of the jet are captured.

In the following subsections, we analyze the general features of time-averaged images and their spin dependence. As part of further analyses, we also discuss the statistical deviation of jet widths in subsections \ref{subsec:acceleration} and examine the jet acceleration profile using the jet width profiles in \ref{subsec:deviation}, as well as calculating the correlation between the jet power $P_{\rm jet}$ and the jet width in \ref{subsec:power-width}.

\subsection{Time-averaged Images, Magnetic Fields and Plasma Bulk Motion}\label{subsec:average}


In the following subsections, we analyze the features of time-averaged images at 86GHz. We have confirmed that the image features are essentially the same at 230GHz and 345~GHz due to the low optical depths at these frequencies for self-absorption and Faraday effects in the jet, although the intensity is lower.

\subsubsection{Total intensity images}\label{subsubsec:total}

\begin{figure*}
\begin{center}
	\includegraphics[width=19cm]{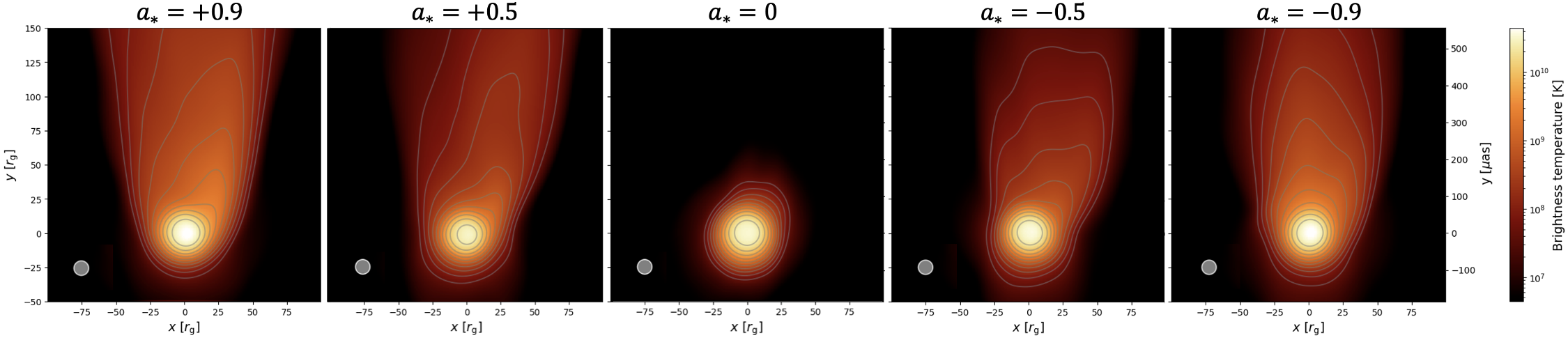}
\end{center}
    \caption{
    Time-averaged total intensity images for the five selected models ($a_*=+0.9$, +0.5, 0, -0.5, -0.9), averaged over $5000~t_{\rm g}$.
    }
    \label{fig:average_total}
\end{figure*}

\begin{figure*}
\begin{center}
	\includegraphics[width=19cm]{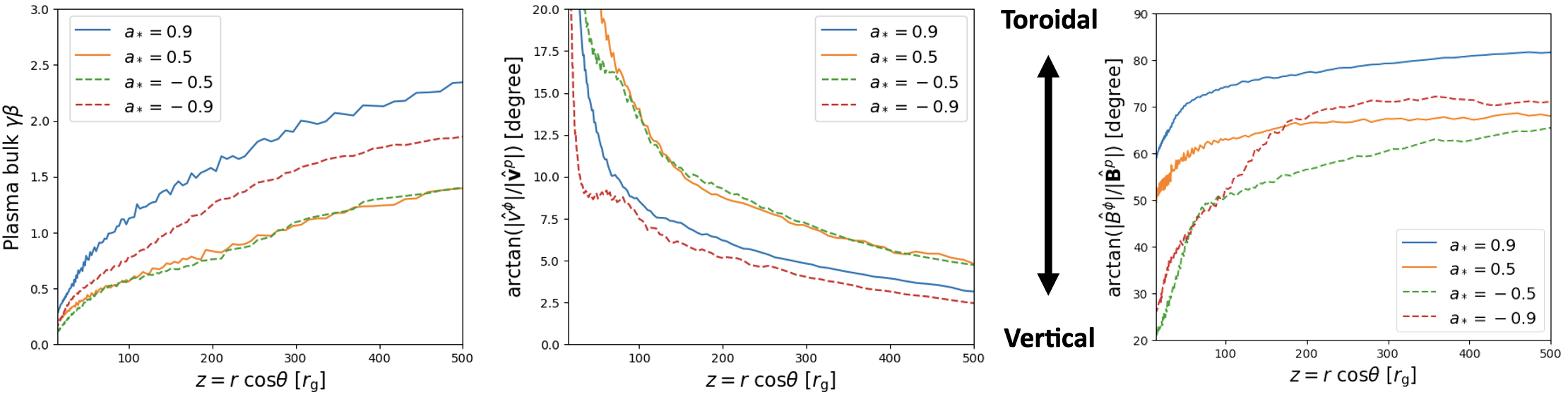}
\end{center}
    \caption{
    Profiles vs $z$ of the plasma bulk velocity multiplied by the Lorentz factor $\gamma\beta$ (left), the helicity angle of plasma bulk motion  $|\hat{v}^\phi|/|\hat{\bm v}^p|$ (middle) and of magnetic fields in the ZAMO frame $|\hat{B}^\phi|/|\hat{\bm B}^p|$ (right) in four GRMHD models with spinning BHs. Each quantity is  averaged over a duration $5000~t_{\rm g}$ and over azimuthal angle $[0, 2\pi)$. The highly magnetized jet region with $\sigma \in [5,10]$ is considered in calculating the profiles. Note that the scales of the vertical axis are different in the central and right panels.
    }
    \label{fig:velocity_helicity_profile}
\end{figure*}

Here, we examine the magnetic field configuration and relativistic plasma bulk motion that can be extracted from images of the jet launching region. 
The time-averaged total intensity images over $5000~t_{\rm g}$ for the five spin cases are shown in Fig.~\ref{fig:average_total}. 

The images overall successfully produce an extended approaching jet, except in the nonspinning case. As shown in \cite{2022MNRAS.511.3795N}, the $a_* = 0$ (Schwarzschild) BH does not produce highly magnetized relativistic plasma in the outer region, or if it does exist, it is only temporary. As a result, this model cannot produce a bright jet through the relativistic Doppler-beaming effect, unlike in the other cases. It is notable that our non-thermal eDF prescription naturally produces this effect.

Comparing the spinning cases, we find that the moderate spin cases, $a_* = \pm0.5$, produce a less vertically-extended and more asymmetric jet than the fast spin cases, $a_* = \pm0.9$. This can be explained by the Doppler-beaming effect due to the helical bulk motion of the plasma. Since the present 86 GHz image shows a relatively symmetric jet (\citealp{2023Natur.616..686L}), this itself may be an important spin constraint \rev{(though it should be ultimately discussed bearing the observed limb-brightened feature in mind, as mentioned in subsection \ref{subsec:nonthermal})}. 
The robustness of this signal with respect to the non-thermal eDF prescription must be investigated in future work, but we find that the signal originates from the fluid itself (GRMHD as opposed to GRRT).

In the left and central panels of Fig.~\ref{fig:velocity_helicity_profile}, the time- and azimuthal-averaged GRMHD profiles of the plasma bulk velocity mulitplied by the Lorentz factor $\gamma \beta$ and its helicity angle ${\rm arctan}(|\hat{v}^\phi|/|\hat{\mathbf{v}}^p|)$ are plotted along the $z$-axis, respectively. Here $\gamma = \alpha u^t$, $\alpha$ is the lapse,  and $\beta = \sqrt{1-\gamma^{-2}}$ (thus $\gamma \beta = \sqrt{\gamma^2-1}$); $\hat{v}^\phi = \sqrt{g_{\phi\phi}}{u}^\phi/u^t$ and $\hat{\mathbf{v}}^p = (\sqrt{g_{rr}}{u}^r\hat{\mathbf{r}} + \sqrt{g_{\theta\theta}}{u}^\theta\hat{\boldsymbol{\theta}})/u^t$ are the toroidal and poloidal components of the velocity field, respectively; and the profiles are calculated for the jet region with $\sigma_{\rm m} \in [5,10]$. 

In the left panel in Fig.~\ref{fig:average_total}, we see that the spinning BHs show persistent acceleration of plasma up to the relativistic regime $\gamma\beta > 1$ within $500~r_{\rm g}$ along the jet.\footnote{$\gamma\beta = 1$ and $2$ correspond to $\beta \sim 0.71c$ and $\sim 0.89c$, respectively.} 
The profiles also show spin dependence, where higher spins result in stronger plasma acceleration compared to lower spins, both in the prograde and retrograde cases. This directly explains the less vertically-extended jets in the lower spin cases, due to weaker Doppler beaming.

The central panel exhibits vertical plasma velocity overall, but more helical motion for the lower spins. 
The helical motion of the plasma results in a gap in the beaming effect between the two sides of the jet in the image. 
In the images shown in Fig.~\ref{fig:average_total}, the right side of the jet approaches the observer and gains more beaming, while the left side recedes and experiences less beaming. 
This tendency is amplified with more helical plasma flow, leading to greater asymmetry for lower spin values.
Since the helicity angles are close to the observer's viewing angle of $i = 163\degr$ ($17\degr$), even a small difference in the angle results in a significant difference in the images. 

This explains one edge- or spine-brightened features in the jet-launching region in the time-averaged total intensity images. 
In subsection \ref{subsec:nonthermal}, we compare these features with existing observations based on the nonthermal electron prescription.


\subsubsection{Linear polarization images}\label{subsubsec:LP}


\begin{figure*}
\begin{center}
	\includegraphics[width=19cm]{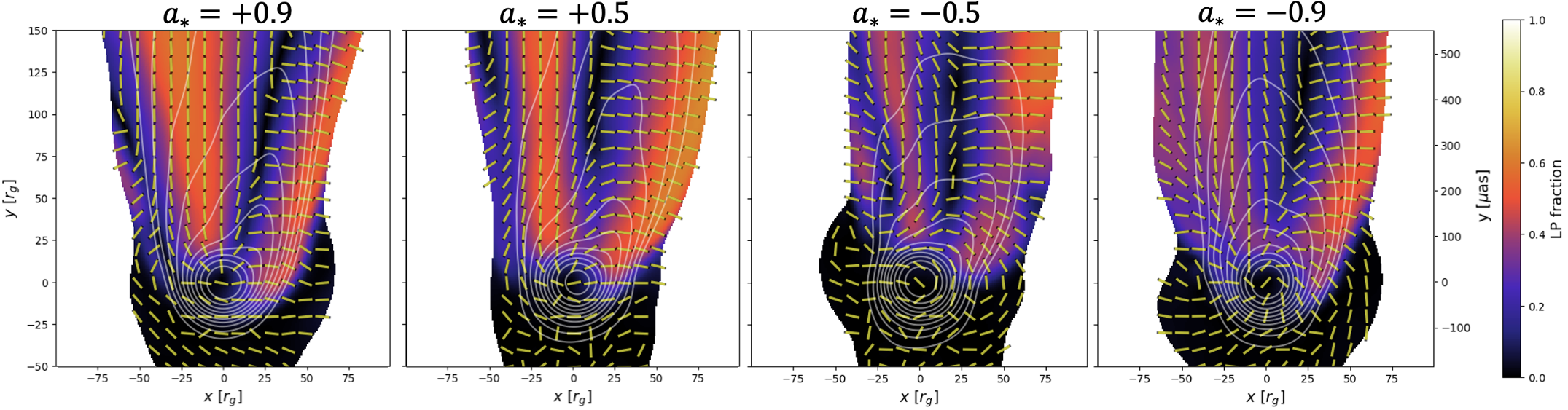}
\end{center}
    \caption{
    Linear polarization (LP) maps for four models with spinning BHs, averaged over $5000~t_{\rm g}$. 
    The line and color contours denote the total intensity (same as in Fig.~\ref{fig:average_total}) and the LP fraction, respectively. The LP vector EVPAs are overwritten as ticks.
    }
    \label{fig:average_LP}
\end{figure*}

\begin{figure*}
\begin{center}
	\includegraphics[height=6cm]{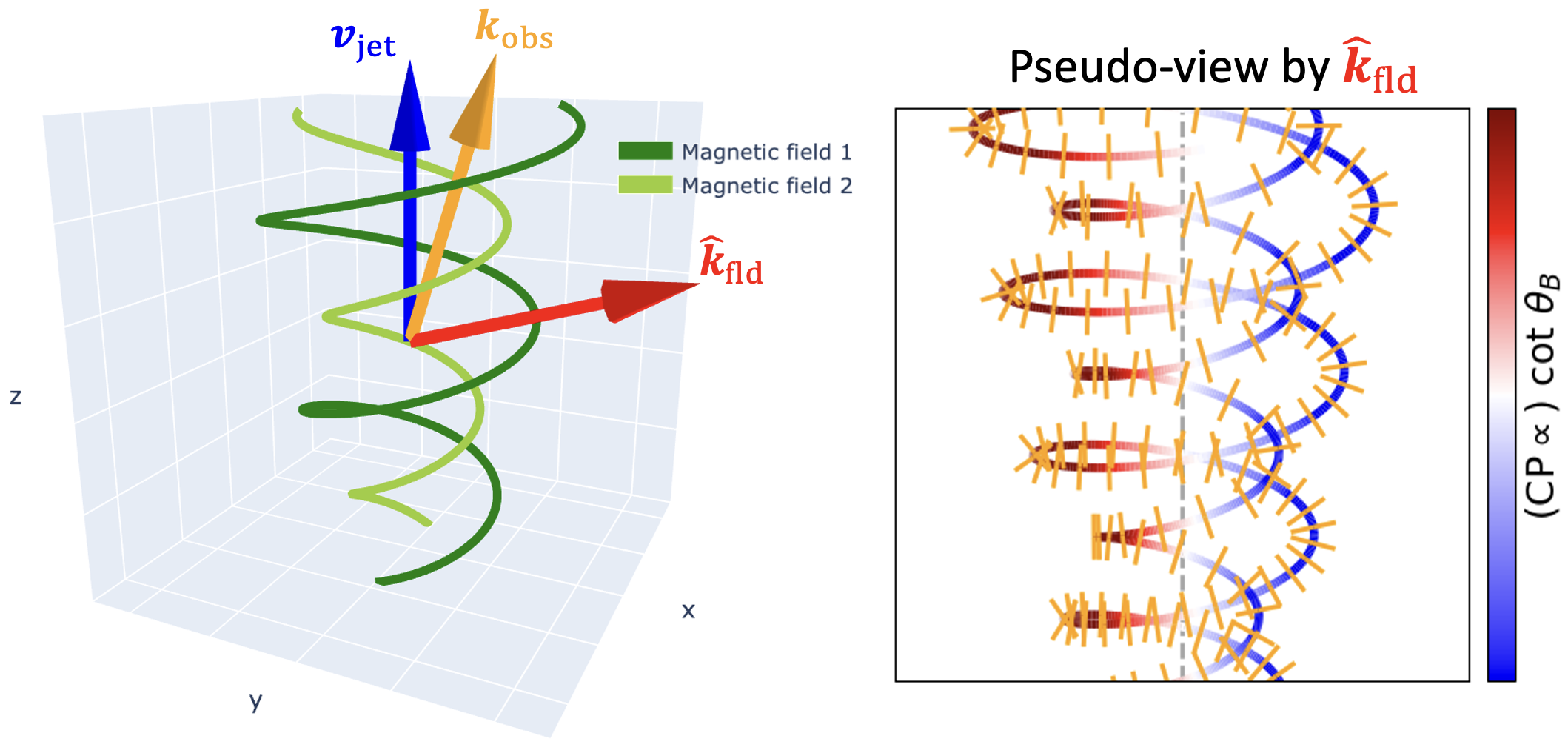}
\end{center}
    \caption{
    Left: schematic picture of helical magnetic field lines 1 (outer, green) and 2 (inner, lightgreen) and the relativistic aberration between the light propagation vector in the observer frame, $\mathbf{k}_{\rm obs}$ (orange arrow), and that in the fluid rest frame, $\hat{\mathbf{k}}_{\rm fld}$ (red arrow), by bulk velocity of jet $\mathbf{v}_{\rm jet}$ (blue arrow). Here the bulk motion is assumed to be purely vertical. 
    Right: pseudo-view of the two field lines and their polarization as viewed by the aberrated $\hat{\mathbf{k}}_{\rm fld}$. 
    The two field lines are overplotted with LP vector ticks.
    The red-blue contour of the lines describes ${\rm cot}~\theta_B$, to which the CP components are proportional (see Appendix \ref{apdx:CP} for CP images). Here $\theta_B$ is the angle between the light propagation vector and mthe agnetic field line. 
    Note that these are the projection of polarization components obtained  in the fluid rest frame onto the pseudo-screen for $\hat{\mathbf{k}}_{\rm fld}$. We actually observe these on the projection onto the observer's screen for $\mathbf{k}_{\rm obs}$ after transforming them into the observer's frame. 
    }
    \label{fig:picture_rel}
\end{figure*}

Time-averaged linear polarization (LP) maps are shown for the four spinning models in Fig.~\ref{fig:average_LP}. Overall, they show an asymmetric pattern of horizontal LP vectors in the right edge (beaming side) and vertical in the center (spine) and left edge (de-beamed side). We also see a horizontal pattern in a narrow region near the left edge. In addition, the counter jet and disk show low LP fractions close to zero due to strong Faraday rotation and depolarization in the low electron temperature ($R_{\rm high} = 160$) disk gas near the BH  (\citealp{2017MNRAS.468.2214M,2020MNRAS.498.5468R,2022ApJ...931...25T}). 

The asymmetric LP vector pattern in the jet does not straightforwardly imply an asymmetry in the magnetic field configuration, since the combination of axi-symmetric helical fields and inclination of observer's viewing angle can cause an asymmetric pattern (see, for example, \citealp{2011ApJ...737...42P,2013MNRAS.430.1504M}). 
In the right panel of Fig.~\ref{fig:velocity_helicity_profile},  we show the helicity angle of the magnetic field in the ZAMO (zero-angular momentum observer) frame, ${\rm arctan}(|\hat{B}^\phi|/|\hat{\mathbf{B}}^p|)$, for the four models. Here, $\hat{B}^\phi$ and $\hat{\mathbf{B}}^p$ are defined in the same way as $\hat{v}^\phi$ and $\hat{\mathbf{v}}^p$, respectively. 
As a whole, the magnetic fields are in a helical configuration.

The key factor to interpret the LP vector patterns in Fig.~\ref{fig:average_LP} is a combination of the relativistic aberration effect and helical magnetic field configuration. 
In the left panel of Fig.~\ref{fig:picture_rel}, we picture two helical magnetic field lines in the rest frame of the plasma which is moving in the $z$ direction with velocity $\mathbf{v}_{\rm jet}$. 
The radiation observed at a certain viewing angle in the observer's frame, which is now $i = 163\degr$ (or $17\degr$) and denoted by $\mathbf{k}_{\rm obs}$ in the picture, is emitted at a larger angle in the plasma rest frame, denoted by $\hat{\mathbf{k}}_{\rm fld}$. 

If the magnetic fields were projected onto this pseudo-screen for $\hat{\mathbf{k}}_{\rm fld}$, they would form vertically compressed ``$\alpha$'', or ``c $+$ x'' shapes as in the right panel of Fig.~\ref{fig:picture_rel}. The LP vectors and CP components in the fluid rest frame are denoted by the ticks and color contour of line, respectively. 
We obtain an ordered vertical LP vector pattern in the broad region in the left side, reflecting the pattern on compressed ``c'' shapes in the central panel, and a horizontal pattern in the edge region of the right side, reflecting that on the right side of the ``x''. 
(Though there are horizontal vectors in the left edge of ``c'' shapes, they are limited to narrow region and subordinate to the dominant vertical vectors.)
In the inside region of the right side, around the cross points of ``x'', the vectors are disordered and cancel each other out. 

Here, we should note that these polarization components are actually observed on the observer's screen for $\mathbf{k}_{\rm obs}$, after transformed into the observer's rest frame (see \citealp{2003A&A...403..805P,2003ApJ...597..998L}; see also \citealp{1979ApJ...232...34B} for the relativistic {\it swing} effect). 
Nevertheless, the larger viewing angle due to the aberration here induces a more asymmetric pattern in the images in Fig.~\ref{fig:average_LP}.
In this way, the relativistic jet with helical magnetic fields gives an asymmetric pattern of LP vectors horizontal in the right side and vertical in the spine and left side of the jet. 
In the transition region between the two patterns, the LP vectors are inclined and the LP fraction is low (see subsection \ref{subsec:LPCPspindep}; see also \citealp{2023ApJ...959L...3D} for depolarization in shear layer.)



For comparison, we show the images and schematic picture in which the relativistic effects are turned off, in Appendix \ref{apdx:nonrel}. In this case, $\hat{\mathbf{k}}_{\rm fld}$ coincides with $\mathbf{k}_{\rm obs}$ as pictured in Fig.~\ref{fig:picture_nonrel}, showing a symmetrical pattern which reflects the projection of helical magnetic field lines from the face-on-like view in Fig.~\ref{fig:nobeaming}.

\bigskip

So far, we observed that relativistic aberration generally produces an asymmetric LP vector pattern in the images of the jet-launching region. In subsection \ref{subsec:LPCPspindep}, we discuss the BH spin dependence of the asymmetry in the LP images, focusing on the magnetic field helicity and jet acceleration profiles. In addition, we also note a symmetrical bipolar CP component pattern in the jet images and its spin-depenedence in Appendix \ref{apdx:CP}.


\section{Discussion}\label{sec:discussion}

\subsection{Estimate of Jet Acceleration from Observables}\label{subsec:acceleration}


\begin{figure*}
\begin{center}
	\includegraphics[width=19cm]{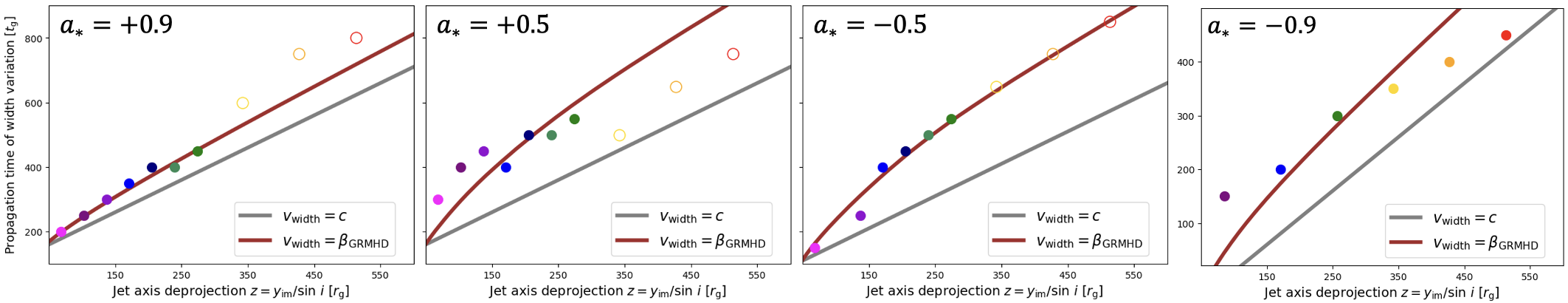}
\end{center}
    \caption{
    Profile of the propagation time of changes in the jet width along the jet axis for the spinning BH models. 
    The vertical positions of the colored dots show $\Delta t_{\rm peak}$, which corresponds to the time lag of the peak of the correlation functions between the total intensity flux and the jet width at each $y$, i.e., the time gap between when a fluctuation in the total flux occurs and when its effect is seen in the jet width profiles in Figs.~\ref{fig:profiles}, \ref{fig:a09}, \ref{fig:a05}, \& \ref{fig:am05} (see also Appendix \ref{apdx:a09} for the other spin models than $a_* = -0.9$). 
    The horizontal axes correspond to the $y$ coordinate on the image deprojected onto the $z$-axis in the fluid model coordinates by the observer's viewing angle $i = 163\degr$ ($17\degr$). The black line corresponds to propagation at speed $c$. The brown line is the predicted profile if changes in the jet width propagate at the average plasma bulk velocity in the GRMHD model $\beta_{\rm GRMHD}$, i.e., by integrating $dt = dz/\beta_{\rm GRMHD}$ from $z = 40~r_{\rm g}$ to each $z$. Here, the plasma bulk velocity $\beta_{\rm GRMHD}$ is obtained as in the left panel of  Fig.~\ref{fig:velocity_helicity_profile}. 
    In the rightmost panel, an offset of $-50~t_{\rm g}$ is added to the two profiles since the total intensity flux is delayed by $50~t_{\rm g}$ relative to $\phi_{\rm BH}$, as shown in the left panel of Fig.~\ref{fig:correlation}. 
    The same applies to the other panels, based on the $\phi_{\rm BH}$-total flux offset in each model.
    }
    \label{fig:acceleration}
\end{figure*}

In subsection \ref{subsec:correlation}, we showed that the jet widths in the image and the total intensity flux are strongly correlated with the normalized magnetic flux at the event horizon $\phi_{\rm BH}$, with the time delay increasing monotonically with distance. 
This result implies that the change in jet widths propagates from the BH to the jet, and motivates us to explore whether it might be possible to estimate the plasma acceleration profile by observing variations in the jet width. 

We calculate the correlation function between the total flux and the jet width at each $y$ on the image for the spinning BH models (the black and green profiles in the right panel of Fig.~\ref{fig:profiles} for the $a_* = -0.9$ model and in the leftmost panels of Figs.~\ref{fig:a09}, \ref{fig:a05}, \& \ref{fig:am05} for the other spin models). The delay $\Delta t_{\rm peak}$ as a function of the deprojected distance along the jet axis, $z = y / {\rm sin}~i$, is plotted in Fig.~\ref{fig:acceleration}. The delays correspond to the time it takes for changes in the jet width to propagate along the axis, up to each $y$ position in the image. 

We also plot two model curves for comparison.
The straight gray line corresponds to the jet width variation propagating at a constant speed of $v=c$. The brown line assumes that the propagation speed is the average plasma bulk velocity in the GRMHD model, i.e.,
we integrate $dt = dz/\beta$ (and $dt = dz/c$ for the $v=c$ line) from $z = 40~r_{\rm g}$ to each $z$, where the plasma bulk velocity $\beta$ is obtained as in the left panel of  Fig.~\ref{fig:velocity_helicity_profile}. 
We here add an offset of $\Delta t_{\rm peak}$ of the $\phi_{\rm BH}$-total flux correlation to the two profiles, since the total intensity flux is delayed (or precedes) relative to $\phi_{\rm BH}$, as shown in the left panel of  Fig.~\ref{fig:correlation}.\footnote{Here, we assume that the change in jet shape begins to propagate downstream at the moment the change in $\phi_{\rm BH}$ occurs.} 

Figure~\ref{fig:acceleration} suggests that there is acceleration up to the relativistic regime, \rev{70 to 90 percent of the speed of light}, for all the spinning BH models. 
They also persistently show reasonable agreement with the average bulk velocity from GRMHD up to $z \sim 300~r_{\rm g}$. 
These results suggest that it may be possible to measure the jet acceleration profile from jet width measurements. 
Note that the times shown in the plots correspond to coordinate time in the simulation frame, while the {\it observed} time lags will be shorter because of relativistic Doppler beaming. This has  not been included in the calculations presented here.


\subsection{Variability of Jet width and $\Phi_{\rm BH}$}\label{subsec:deviation}



\begin{figure*}
\begin{center}
	\includegraphics[width=10cm]{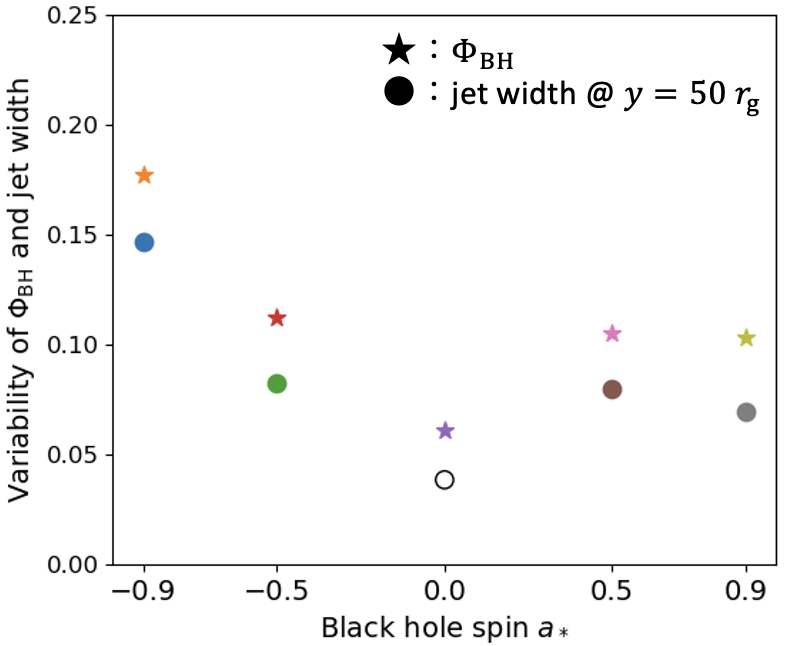}
\end{center}
    \caption{
    The variability $\sigma / \mu$ (rms fluctuations divided by the mean) of the non-normalized magnetic flux on the event horizon $\Phi_{\rm BH}$ (denoted by the stars) and jet widths at $y = 50~r_{\rm g}$ on the images (circles) for the five models. Note that the jet width for $a_* = 0$ (empty circle) is exceptionally measured at $y = 10~r_{\rm g}$. 
    }
    \label{fig:deviation}
\end{figure*}

\cite{2022MNRAS.511.3795N} pointed out that the variability of the {\it unnormalized} magnetic flux on the event horizon, $\Phi_{\rm BH}$, increases with the BH spin value for retrograde models, while it remains almost constant for prograde models. 

To check the relationship between the variability in GRMHD models and GRRT images, we calculate the variability of jet width at $y = 50~r_{\rm g}$ on the image for the five models. 
Here, the variability is calculated for each duration of $1000~t_{\rm g}$, dividing the snapshots into five bins, and is then averaged over the bins, following \cite{2022MNRAS.511.3795N}. 

Fig.~\ref{fig:deviation} shows the variability of $\Phi_{\rm BH}$ and jet width. 
First, our $5000~t_{\rm g}$ duration data successfully reproduce the $\Phi_{\rm BH}$ variability shown in \cite{2022MNRAS.511.3795N} for the longer duration of $50000~t_{\rm g}$. 
Second, the variability of the jet width shows similar values to the variability in $\Phi_{\rm BH}$, preserving the dependence on BH spin. 
The slight decrease in the magnitude of the variability can be attributed to the mock-observational blurring in the images, which tends to smooth out subtle changes in the jet shape. 

We also find that the variability in the jet width reproduces the behavior of $\Phi_{\rm BH}$ rather than $\phi_{\rm BH}$. 
The latter shows variability increasing with the spin value for both retrograde and prograde models \citep{2022MNRAS.511.3795N}. 
Furthermore, we also confirmed that the jet widths show slightly better correlation with $\Phi_{\rm BH}$ than $\phi_{\rm BH}$, while they yield  insignificant correlation with the mass accretion rate $\dot{M}$. 
Based on this, we can say that the jet widths selectively reflect the dynamics of magnetic flux on the horizon, not the mass accretion process onto the BH.

In this way, we saw that the jet width gives the variability monotonically increasing with BH spins for the retrograde models, and the almost unchanging variability for the prograde models.
This tendency is consistent with the variability of the absolute magnetic flux $\Phi_{\rm BH}$. 
This result suggests that we can expect to determine the handedness and value of BH spin through long-term observations of the jet-launching region. 

\subsection{Jet power-width correlation}\label{subsec:power-width}


The result in the last subsection leads us to another important quantity in the BZ process: the jet power. 
In the right panel of Fig.~\ref{fig:correlation}, we show the correlation functions between the jet power $P_{\rm jet}$ (see the bottom panel of Fig.~\ref{fig:snapshots} for the variability profile) and the jet width at each $y$ position on the image for the $a_* = -0.9$ model. 
\rev{
All the jet widths exhibit strong correlations with the jet power which are comparable with those with $\phi_{\rm BH}$. 
}

These results suggest that the jet widths are also a good indicator of jet power. Thus, we can expect to survey the jet power and efficiency, which serves as a litmus test of the BZ process. 
In addition, since the jet power derived from the rotational energy of the BH results in spin-down (\citealp{2022MNRAS.511.3795N,2023ApJ...954L..22R,2024arXiv241007477R}), the observation of jet width can provide a foothold for investigating the BH spin evolution.

\subsection{Spin-dependence of the LP maps}\label{subsec:LPCPspindep}


The LP maps in Fig.~\ref{fig:average_LP} show a spin-dependence, and we noted their general asymmetric appearance in sub-subsection \ref{subsubsec:LP}. 
Lower spins show more asymmetric patterns than the higher ones, as seen in the difference in the location of the zero LP-fraction line which demarcates the horizontal pattern in the right edge and the vertical pattern in the middle. 


This is triggered by the spin-dependence of the helicity angle of the magnetic field.  As shown in the right panel of Fig.~\ref{fig:velocity_helicity_profile}, the lower spin cases consistently show a less toroidal (more vertically-dominated) field structure compared to the higher spins, which wind up the field lines more strongly. 

If the plasma bulk motion is the same, this results in more compressed ``c $+$ x'' shapes, or ``$<$'' shapes of projected magnetic field lines, as shown in Fig.~\ref{fig:picture_lesstor} in Appendix \ref{apdx:CP}. 
In such a case, the border of vertical and horizontal patterns is shifted to the left side with the cross point of ``x'', as also shown in \cite{2013MNRAS.430.1504M}. 
Therefore, the lower spins give more asymmetric pattern with the horizontal LP vector filling larger area in the right side of jet. 

\cite{2021A&A...656A.143K} calculated images based on relativistic MHD models with three kinds of magnetic field configuration. Their Helical model exhibits a horizontal vector pattern in the beamed side of the jet, while their Toroidal shows a symmetrical pattern of vertical vectors in the spine and horizontal vectors in the two edges, as in our non-relativistic image in Fig.~\ref{fig:nobeaming}. 
Our images in Fig.~\ref{fig:average_LP} are in between their models, showing the asymmetric pattern which reflects the helicity of magnetic fields from the relativistically-aberrated view. 

In addition, our LP maps also show inclined vectors in the transition between horizontal and vertical patterns. They indicate whether the radiation comes from the front or back side of the jet (though the polarization fraction is low). 
As shown in Figs.~\ref{fig:picture_rel} and \ref{fig:picture_lesstor}, the ``/'' ticks come from the back of the jet while the ``\textbackslash'' ticks come from the front. 
These features can be a tracer of jet morphology along the axis, if detected.

\bigskip

These results in the LP maps suggest a new method for investigating the magnetic field configuration, the plasma acceleration profile, and the BH spin. 

Recently, \cite{2024arXiv241000954G} surveyed the LP maps using semi-analytical force-free jet models. 
Our and their results are qualitatively consistent in the tendency that the moderate spin $|a_*|=0.5$ model tends to exhibit more asymmetric pattern in the inner and outer jet than the high spin $|a_*|=0.9$ model, reflecting less toroidal magnetic field configuration due to larger light cylinder radius (Z.~Gelles, private communication). 

Meanwhile, they also pointed out that the LP vector pattern switches from radial to azimuthal at the light cylinder radius, where the plasma bulk velocity becomes relativistic and starts to suppress the magnetic field component parallel to the motion. 
In contrast, our LP images do not show such transition due to moderate acceleration. As in the left panel of  Fig.~\ref{fig:velocity_helicity_profile}, plasma bulk motion in the region with $\sigma < 10$ barely reaches $\gamma \beta \sim 1.2$ in the $a_* = 0.9$ model and $\sim 0.5-0.7$ in the other spins at $z = 100~r_{\rm g}$, while their force-free models yield $\sim 3$. 
Addressing this gap between the force-free and GRMHD models is an important task for future studies.

\bigskip

In observational approaches, \cite{2009MNRAS.393..429O} reported detection of the vertical and horizontal LP vector patterns along active galactic nucleus jets at centimeter wavelengths on tens-parsec scale. 
\cite{2018ApJ...855..128W} detected a persistent LP vector pattern on sub-mas scale in the M87 jet, implying the presence of significant toroidal magnetic field components.
Upcoming observations of M87* with high angular resolution and high sensitivity will shed new light on the black hole jet mechanism.



\subsection{Comparison with observations and non-thermal contribution}\label{subsec:nonthermal}


In subsection \ref{subsubsec:total}, we saw single-edge or spine-brightened jets in time-averaged images. Meanwhile, observations of the base region of M87's jet at 86~GHz show a double-edged, limb-brightened morphology (e.g., \citealp{2018A&A...616A.188K,2023Natur.616..686L}).
There is a large gap by a factor of $\sim 5$ between the intensities in the edges and the spine at $y =$ 0.25~mas ($\approx 68~r_{\rm g}$ on our image) along the jet (\citealp{2024arXiv240900540K}).

This discrepancy should be addressed with future work, surveying the nonthermal electron prescription. 
In introducing  nonthermal electrons in the electron energy distribution in this work, we assumed energy partition with the energy in the nonthermal electrons being only 3\% of that in the thermal electrons, as mentioned in subsection \ref{subsec:GRRT}. 
This naturally makes the nonthermal component subordinate to thermal in the electron distribution function (eDF), and is a conservative or passive choice. 
We have confirmed that the contribution of nonthermal emission in the image is overall $\lesssim 10~\%$ in the jet region. 
\cite{2022A&A...660A.107F} calculated GRRT images based on the kappa-eDF (see, for example, \citealp{2016ApJ...822...34P}), finding that a nonthermal dominant prescription in partition with the magnetic energy produced emission over a longer region of the jet. 

\rev{
In addition, the power-law index $p$ is fixed to 2.5 in the whole region in this work. 
\cite{2018ApJ...862...80B} and \cite{2023ApJ...944..122M} suggested that the spectral slope is steeper in weakly magnetized region with $\sigma\le1$. 
Recently, \cite{2024SciA...10N3544Y} started investigating spatially-variable p based on the particle acceleration via magnetic reconnection. In nonthermal-dominant regime, energy prescription should be surveyed bearing in mind various acceleration mechanisms and the synchrotron cooling effect.
}

In addition, there is uncertainty in the sigma cutoff we applied. 
Recently, \cite{2024MNRAS.532.3198C} developed a hybrid model of GRMHD and force-free electrodynamics which avoids the need for a density floor. They pointed out that traditional GRMHD simulations are relatively safe up to $\sigma \approx 25$. 
We found that using a high $\sigma$ cutoff value gives a more extended jet image (as also seen in \citealp{2022A&A...660A.107F}) but less edge-brightened jets. 
It is an urgent task to identify the parameters that reproduce the observational results.
 
\section{Conclusions}\label{sec:conclusions}

In this work, we investigated the variability and large-scale structure of magnetic fields in the jet-launching region up to several hundred gravitational radii $r_g$, through GRRT image calculations based on GRMHD models. 
We focus on the science attainable with improvements to dynamic range, spatial resolution, and temporal resolution, which will be made possible by ngEHT and BHEX.
Our conclusions are summarized as follows; 
\begin{itemize}
    \item {\bf Jet Archaeology and Forecasting}: Variations in the jet width reflect those of the normalized magnetic flux on the event horizon, $\phi_{\rm BH}$, with a time delay. This enables {\it archaeology} of magnetic field dynamics on the black hole through measurements of the extended jet width, as well as {\it forecasting} upcoming changes in the jet morphology from variations in the total intensity flux or high-energy eruptive events. By examining the variability in magnetic flux and jet morphology through these methods, we can expect to verify the MAD model. Note that this paper is focused on the innermost regions of the jet at distances below $10^3r_g$. At larger distances, jet width variations are likely influenced more by fluctuations in the ambient medium rather than magnetic field variability at the black hole \rev{(see, for example, \citealp{2023MNRAS.526.5949N} for helical structures and the Kelvin-Helmholtz instability)}.
    \item {\bf Testing the Blandford-Znajek process}: The jet width variation exhibits acceleration from the upstream to the downstream, matching the velocity profile of the plasma bulk motion. Additionally, the jet widths show a strong correlation with the jet power, $P_{\rm jet}$. The BZ process can be tested by detecting the acceleration profile up to the relativistic regime and measuring the jet power and efficiency.
    \item {\bf Constraining BH Spin}: The variability in jet width shows a dependence on the black hole spin, mirroring that of the magnetic flux, $\Phi_{\rm BH}$. The time-averaged total intensity images display different asymmetric beamed features depending on spin, due to variations in the plasma acceleration profile. Similarly, the time-averaged LP maps show spin-dependent vector patterns, reflecting the helicity of the magnetic field (see also the appendix for spin dependence in CP images).
    Thus, both long-term and high-cadence observations of the jet-launching region may enable novel constraints on the BH spin, independent of those obtained on event horizon scales. 
\end{itemize}

As is typical for these types of studies, there remains significant uncertainty in the prescription of the nonthermal electron distribution function (eDF). The details of the nonthermal prescription is well-known to impact jet morphology on large scales. In future work, it will be essential to explore a variety of prescriptions and aim to reproduce observed features across global scales at multiple frequencies for M87 and other jets. 




\section*{Acknowledgements}

The authors thank Dominic Pesce and Zack Gelles for constructive discussion and comments, and the referee for helpful suggestions. 
This publication is funded in part by the Gordon and Betty Moore Foundation (Grant \#8273.01). It was also made possible through the support of a grant from the John Templeton Foundation (Grant \#62286).  The opinions expressed in this publication are those of the authors and do not necessarily reflect the views of these Foundations. 
YT is grateful for support from JSPS (Japan Society for the Promotion of Science) Overseas Research Fellowships. 
Numerical computations were in part carried out on Cray XC50 at Center for Computational Astrophysics, National Astronomical Observatory of Japan.




\appendix

\section{Magnetic flux-Jet width correlation for different BH spin models}\label{apdx:a09}

\begin{figure*}
\begin{center}
	\includegraphics[width=18cm]{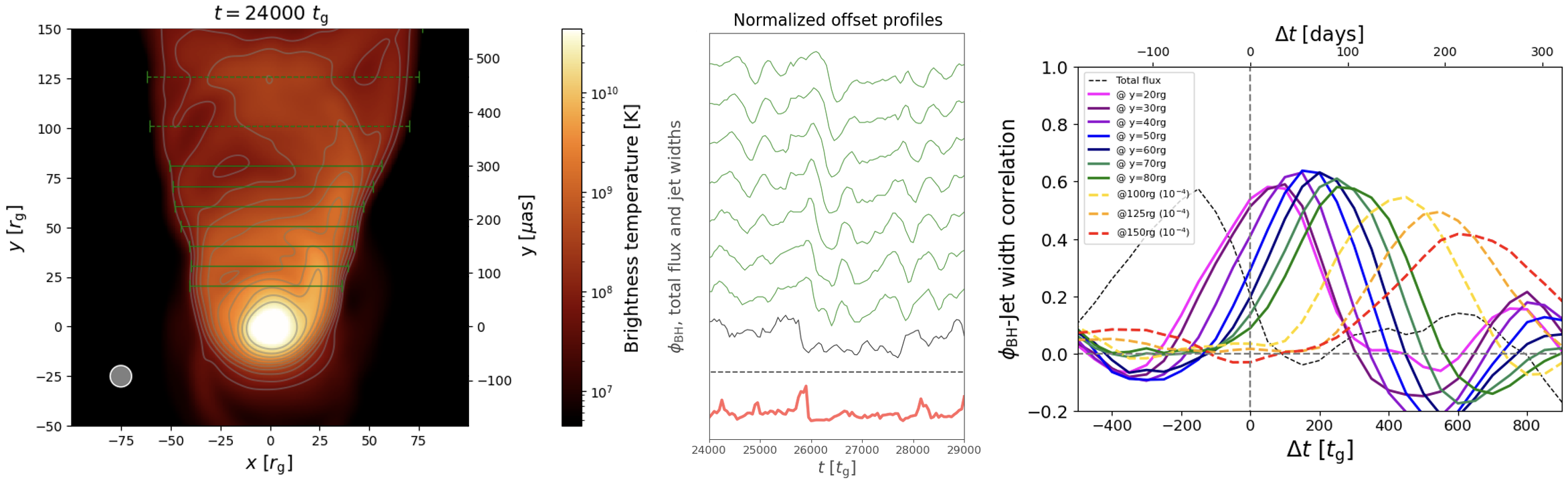}
\end{center}
    \caption{
    Left: a snapshot image for the $a_* = 0.9$ model with measures of jet widths (green lines). For the three dashed lines, we measure the two  edges as the outermost points with $10^{-4}$ times the peak of color contour. 
    Center: Variability profiles of $\phi_{\rm BH}$ (red) and the jet widths (green) defined by the solid lines in the left for the duration of $5000~t_{\rm g}$. The absolute scales of profiles are ignored. 
    Right: Correlation functions between $\phi_{\rm BH}$ and the jet widths (solid and dashed lines), and total intensity flux (dotted line). 
    See \url{https://youtu.be/FwiKspIXjZk} for a movie. 
    }
    \label{fig:a09}
\end{figure*}

\begin{figure*}
\begin{center}
	\includegraphics[width=18cm]{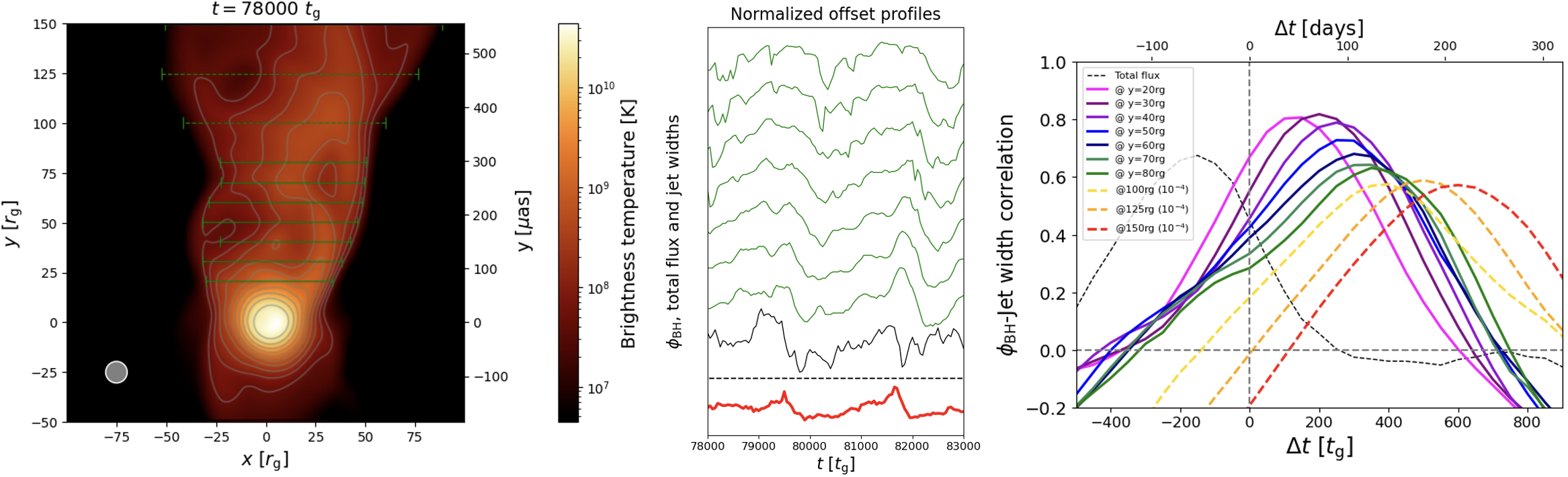}
\end{center}
    \caption{Same as Fig.~\ref{fig:a09}, but for the the $a_* = 0.5$ model.
    }
    \label{fig:a05}
\end{figure*}

\begin{figure*}
\begin{center}
	\includegraphics[width=18cm]{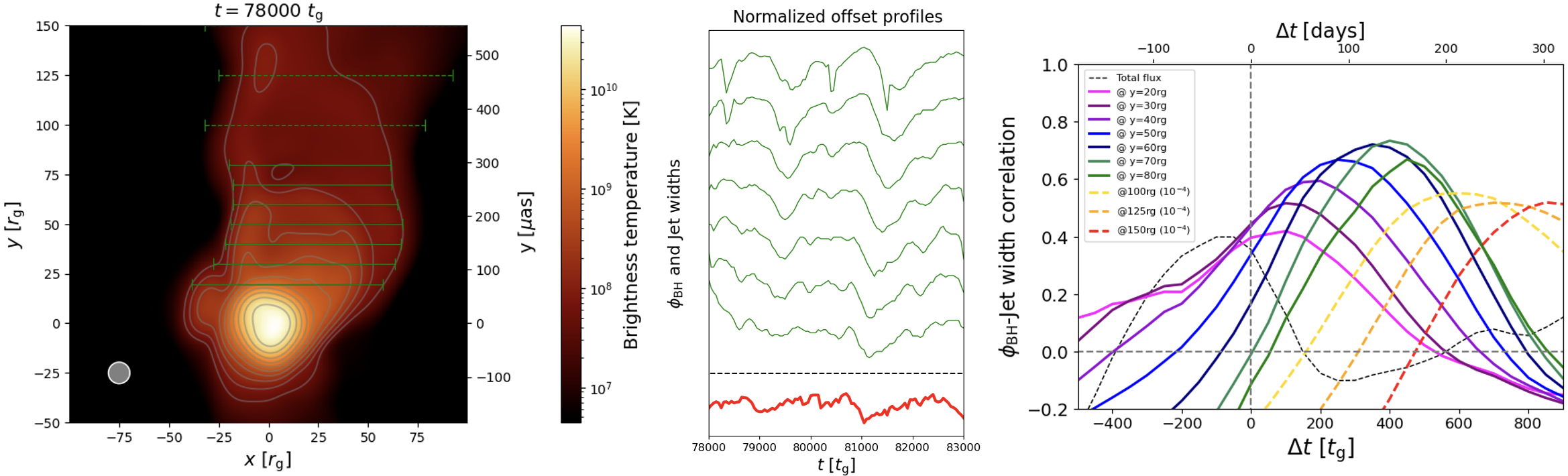}
\end{center}
    \caption{Same as Fig.~\ref{fig:a09}, but for the the $a_* = -0.5$ model.
    }
    \label{fig:am05}
\end{figure*}

A snapshot image with jet width measures, the time profiles of $\phi_{\rm BH}$, total intensity flux, and jet widths, and the correlation function between them are shown in Fig.~\ref{fig:a09} for the $a_* = 0.9$ model, Fig.~\ref{fig:a05} for $a_* = 0.5$, and Fig.~\ref{fig:am05} for $a_* = -0.5$.\footnote{See \url{https://youtu.be/FwiKspIXjZk} for a movie.} 
Here, we measure the jet widths at $20, ..., 80~r_{\rm g}$ (solid lines) at the same threshold of $10^{-3}$ times the peak intensity, as in subsection \ref{subsec:correlation}. 
In addition, the outer jet widths at $100, 125, 150~r_{\rm g}$ (dashed lines) are measured with a lower intensity threshold, $10^{-4}$ times of the peak. This enable to capture the transverse profile of the outer jet, which is fainter especially in the de-beamed side. 


In Fig.~\ref{fig:acceleration} in the main text, the jet acceleration profiles are also shown for the three spin models. 
Here, we commonly calculate $v=c$ and GRMHD-velocity $v=\beta_{\rm GRMHD}$ profiles from $z=40~r_{\rm g}$, and add an offset of the delay between $\phi_{\rm BH}$-total flux in each case. 
The points appear on the GRMHD profile up to $z \sim 300~r_g$ for all the spinning BH models, implying that the inner jet acceleration can be traced by the jet width measurement. 
Furthermore, this result suggest that we can infer the BH spin from the acceleration profile estimated from the jet widths.

\section{\rev{Parameter survey of the jet width correlation}}\label{apdx:survey}

\begin{figure*}
\begin{center}
	\includegraphics[width=10cm]{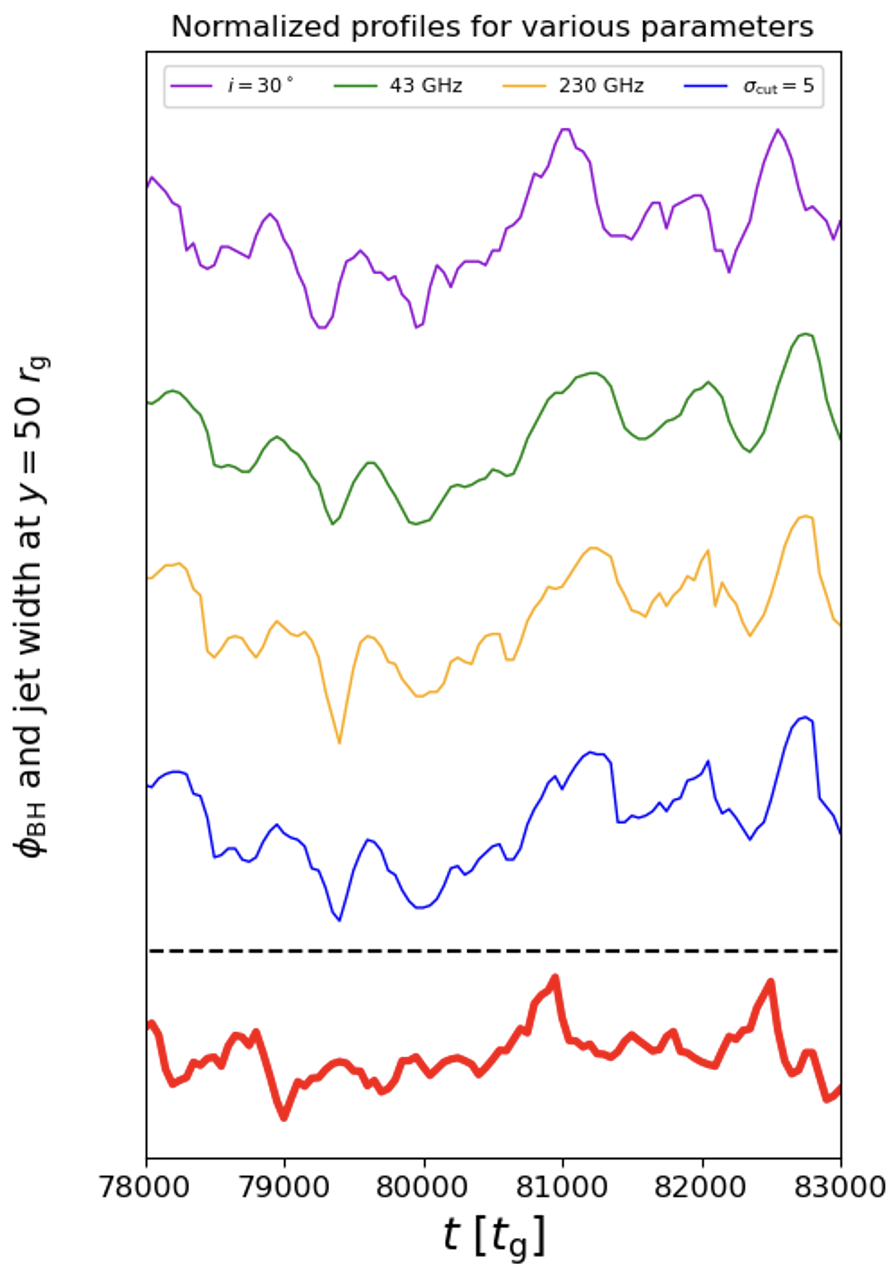}
\end{center}
    \caption{
    \rev{The magnetic flux $\phi_{\rm BH}$ (red) and jet width at $y = 5~r_{\rm g}$ on image for various model parameters; the inclination angle of $30^\circ$ (violet), at 43 (green) and 230~GHz (yellow), and the sigma cutoff of $\sigma \le 5$ (blue). In each case, the other parameters are fixed to the fiducial values for the $a_*=-0.9$ case. The images at 43 and 230~GHz are blurred with a circular Gaussian beam of $80$ and $20~{\rm \mu as}$, respectively.} 
    }
    \label{fig:survey}
\end{figure*}

To check the validity of the results in this work for different model parameters, we show in Fig.~\ref{fig:survey} the profiles of the magnetic flux $\phi_{\rm BH}$ and jet width at $y = 50~r_{\rm g}$ on image for a large inclination angle of $i = 30^\circ$, for the images at 43 and 230~GHz convolved with a circular Gaussian beam of 80 and 20~${\rm \mu as}$, and for a small value in sigma cutoff, $\sigma \le 5$. 
In each case, all the other parameters are set to the fiducial values for the $a_*=-0.9$ case. 

It is confirmed that all the models give a jet width profile reflecting the magnetic flux with a peak correlation coefficient of $\sim 0.8$, comparable with that in the main text. 
Furthermore, we can see that the large inclination case shows a shorter time delay than the other cases. This is because $y = 50~r_{\rm g}$ on image corresponds to a smaller distance from the BH for the larger inclination angle (i.e., $50~r_{\rm g}/{\rm sin}\,30^\circ = 100~r_{\rm g} < 50~r_{\rm g}/{\rm sin}\,17^\circ \approx 170~r_{\rm g}$). 
This implies that the inclination of the jet can be investigated via the jet width analysis.

\section{Images in which the relativistic effects are turned off}\label{apdx:nonrel}

\begin{figure*}
\begin{center}
	\includegraphics[width=16cm]{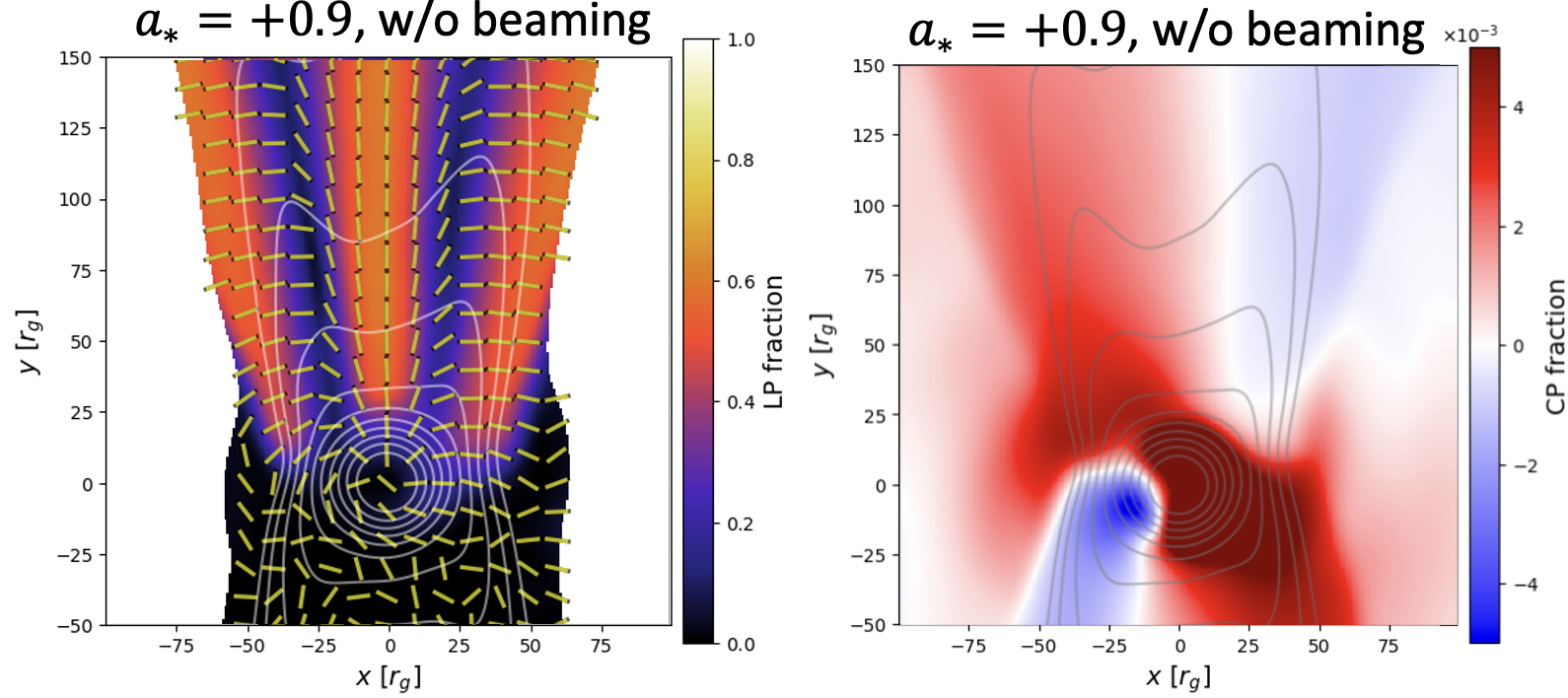}
\end{center}
    \caption{
    The time-averaged LP map (left) and CP image (right) for $a_* = 0.9$, when the relativistic effects of the plasma bulk motion are turned off. 
    Here, the four-velocity $u^\mu$ is set to the ZAMO velocity $u^\mu_{\rm ZAMO}$.
    }
    \label{fig:nobeaming}
\end{figure*}

\begin{figure*}
\begin{center}
	\includegraphics[height=6cm]{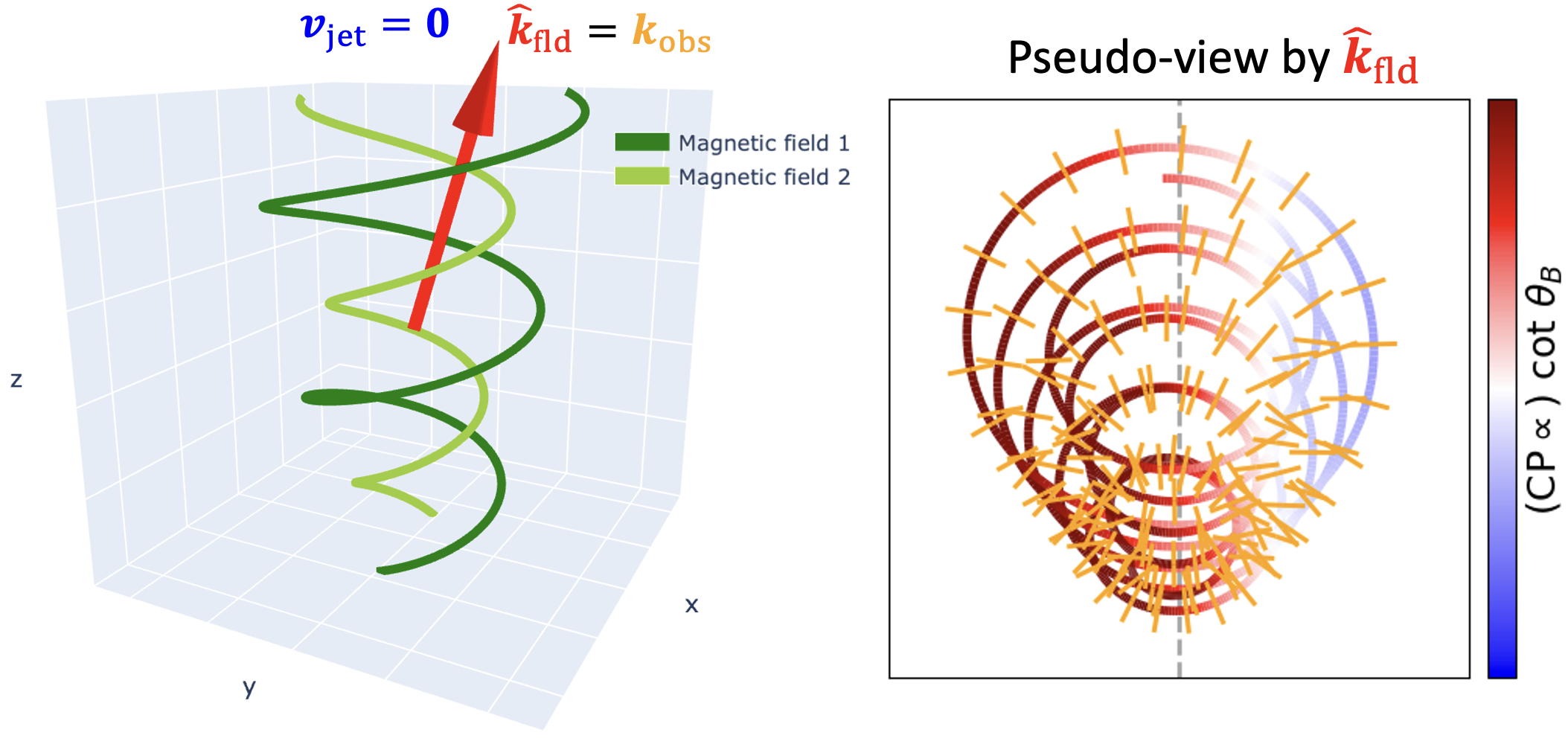}
\end{center}
    \caption{Same as Fig.~\ref{fig:picture_rel}, when the plasma bulk is set to $\mathbf{v}_{\rm jet} = \mathbf{0}$. Here, $\hat{\mathbf{k}}_{\rm fld}$ coincides with $\mathbf{k}_{\rm obs}$, and so does the pseudo-view with the observer's view. 
    Note that the LP vectors are left-right symmetric.
    }
    \label{fig:picture_nonrel}
\end{figure*}

We show the time-averaged LP map and CP image for the $a_* = 0.9$ model in which the relativisitic effects are turned off with the four-velocity set to $u^\mu = u^\mu_{\rm ZAMO}$. Here $u^\mu_{\rm ZAMO}$ is the velocity of ZAMO. 
The picture of helical magnetic field lines and the relativistic aberration effect is also shown in Fig.~\ref{fig:picture_nonrel} for non-relativistic case. 
Here, the LP map shows a symmetrical pattern of horizontal vectors in both edges and vertical vectors in the spine. 
Meanwhile, the CP image presents an asymmetric bipolar pattern leaned to the right (left) side in the approaching (receding) jet.  This clearly demonstrates that relativistic beaming, which is spin dependent, is responsible for the transverse asymmetry in our polarized jet images.

\section{Circular polarization images and spin-dependence}\label{apdx:CP}

\begin{figure*}
\begin{center}
	\includegraphics[width=19cm]{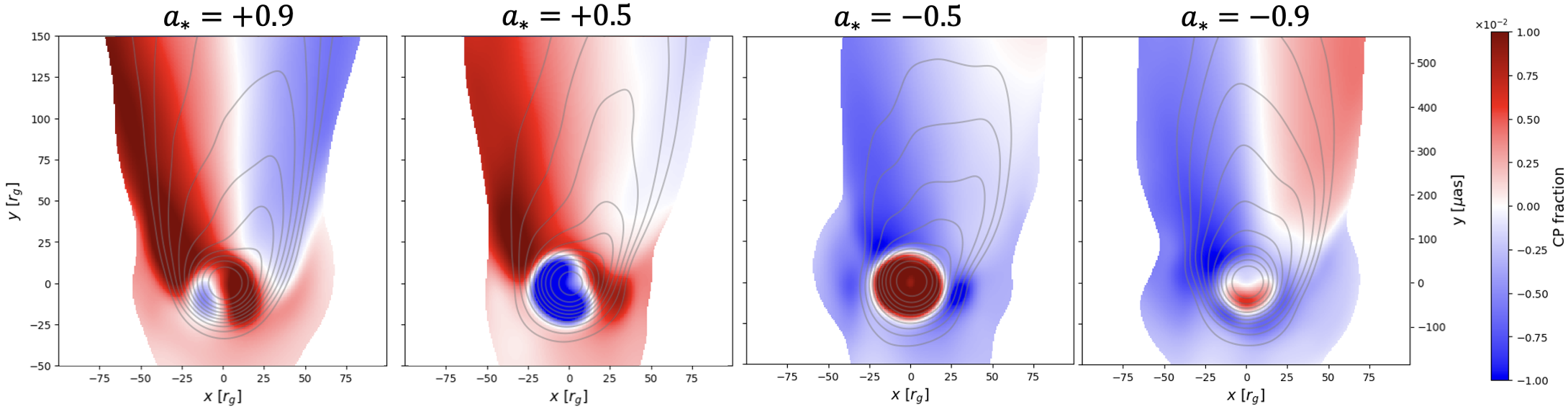}
\end{center}
    \caption{
    The CP maps for five models, averaged over $5000~t_{\rm g}$.
    }
    \label{fig:average_CP}
\end{figure*}

We also investigate time-averaged CP images, shown in Fig.~\ref{fig:average_CP}, but do not find obvious signatures that we believe can be used to infer magnetic flux or spin. 
In the innermost region around the BH within $\sim 20~r_{\rm g}$, they tend to show bright components around the photon ring due to Faraday conversion around the BH (\citealp{2020PASJ...72...32T,2021MNRAS.505..523R,2021MNRAS.508.4282M,2023ApJ...957L..20E,2024ApJ...972..135J}). 
Meanwhile, the conversion is insignificant (less than 0.01 in the optical depth) in the approaching jet. 
Here, the CP components are overall characterized with a bipolar pattern in the horizontal direction. 
That is, the positive components are distributed in the left side, while the negative in the right, for the prograde cases. They are switched for the retrograde cases, reflecting the relation between the polarity of magnetic fields and observer's viewing angle.\footnote{Here, the magnetic field polarity is determined by the initial setting in GRMHD simulation and is common with all the spin cases. Thus, the observer's viewing angle, $i = 163\degr$ (prograde) and $17\degr$ (retrograde), results in the global magnetic fields pointing towards the observer in the prograde case and away from the observer in the retrograde case, respectively.}

This bipolar CP pattern is again interpreted with a combination of helical magnetic fields and relativistic aberration as in Fig.~\ref{fig:picture_rel}. 
It was previously recognized that a helical magnetic field configuration results in a bipolar or quadrant CP pattern on the image for an edge-on view (\citealp{2021MNRAS.505..523R,2021PASJ...73..912T}).
Here, the aberration effect presents an effective edge-on-like view even for a non-edge line of sight, as shown by $\hat{\mathbf{k}}_{\rm fld}$, and produces the bipolar pattern in the images in Fig.~\ref{fig:average_CP}. 

In addition, the CP component is perpendicular to ${\rm cot}~\theta_B$ ($\theta_B$ is the angle between the light propagation vector and the magnetic field vector) in the optically thin case (e.g., \citealp{1977ApJ...214..522J}), giving a large CP fraction for well-aligned or well-antialigned case (i.e, $\theta_B$ close to $0$ or $180\degr$). 
This results in brighter CP components in the left, de-beamed side of the jets in Fig.~\ref{fig:average_CP}.

Meanwhile, the images exhibit a monochromatic CP feature in the counter jet, rather than a bipolar one. On the counter side, the receding plasma bulk motion causes aberration, resulting in a more face-on-like view, in contrast to the approaching jet. As a result, $\hat{\mathbf{k}}_{\rm fld}$ tends to align (or anti-align) with the magnetic field lines in the prograde (retrograde) spin cases, where $0<\theta_B<90\degr$ ($90\degr<\theta_B<180\degr$), producing positive (negative) CP components.

In the right panel of Fig.~\ref{fig:nobeaming}, we show the CP image with relativistic effects turned off. Unlike the LP image, it exhibits a more asymmetric bipolar pattern, as the helical magnetic fields are viewed from a face-on-like angle. Here, the CP components along the vertical midline of the image ($x = 0$) are positive due to the vertical component of the magnetic fields, as illustrated in Fig.\ref{fig:picture_nonrel}. In the absence of relativistic aberration, the counter jet also displays a bipolar pattern. 

\bigskip

\begin{figure*}
\begin{center}
	\includegraphics[height=6cm]{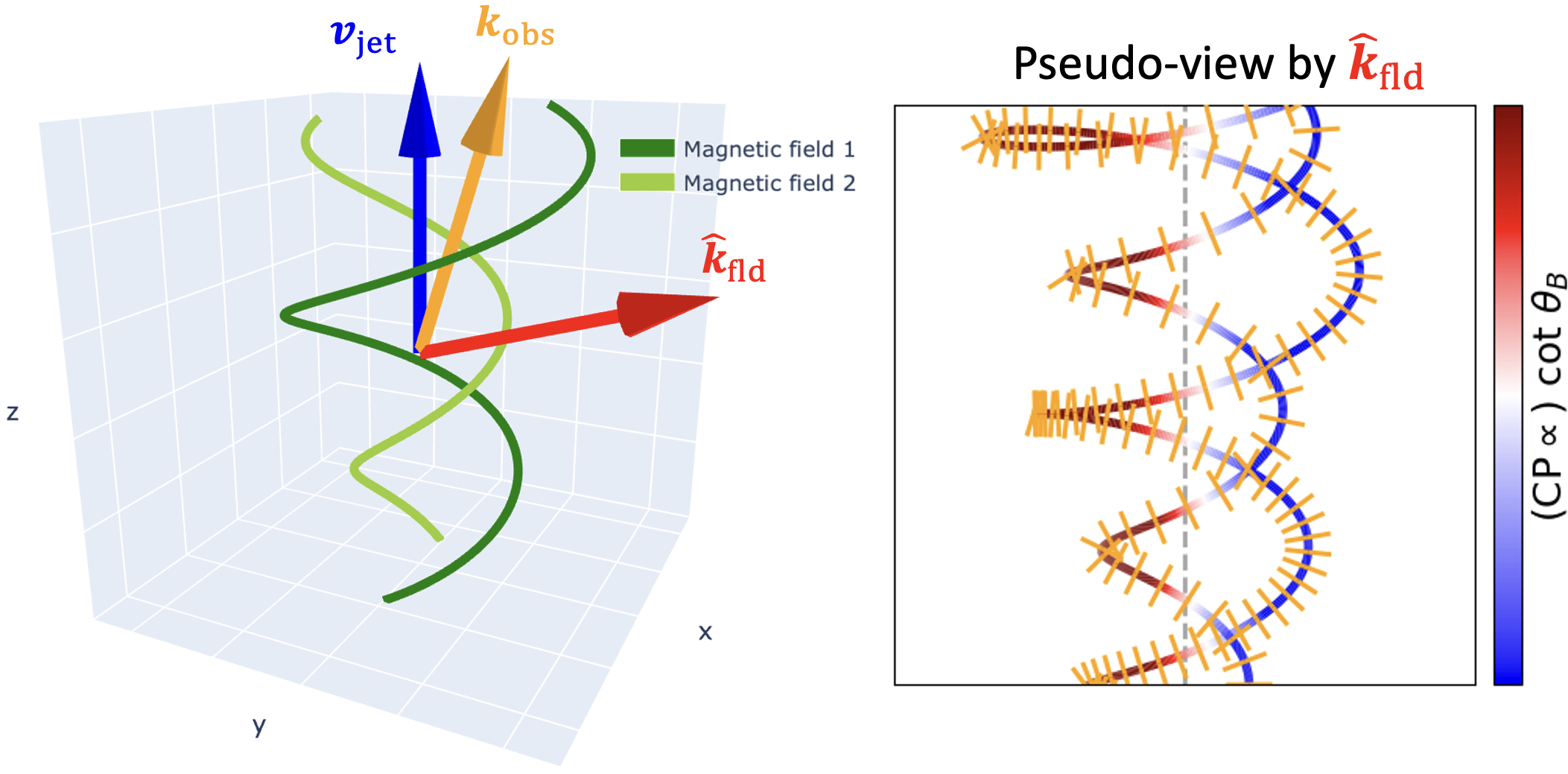}
\end{center}
    \caption{Same as Fig.~\ref{fig:picture_rel}, but for less toroidal magnetic field lines.
    }
    \label{fig:picture_lesstor}
\end{figure*}

The time-averaged CP images in Fig.~\ref{fig:average_CP} also show a tendency to be more asymmetric in the lower spin cases, e.g., the borderline of the bipolar pattern is shifted to the right side for $a_* = \pm0.5$ compared to $a_*=\pm0.9$.
This can also explained by the presence of less toroidal magnetic fields. 
As shown in the pictures in Figs.~\ref{fig:picture_lesstor}, the vertically-dominated field lines weakly enhance the asymmetry of the bipolar CP pattern, causing the border to lean more to the right. 
Thus, they result in more asymmetric CP images in the lower spin cases. 


\newpage

\bibliography{jetwidth_phiBH.bib}
\bibliographystyle{aasjournal}



\end{document}